**Transition Metal-Driven Variations in Structure, Magnetism, and Photocatalysis of Monoclinic $M_3Se_4$ (M = Fe, Co, Ni) Nanoparticles**

Monika Ghalawat[†,‡], Inderjeet Chauhan[†,‡], Dinesh Singh[†,‡], Chinnakonda S. Gopinath[†,‡]*, Pankaj Poddar[†,‡]*

[†]*Physical & Materials Chemistry Division, CSIR-National Chemical Laboratory, Pune 411008, India*

[‡]*Academy of Scientific and Innovative Research (AcSIR), Sector 19, Kamla Nehru Nagar, Ghaziabad, Uttar Pradesh- 201 002, India*

**ABSTRACT**

The transition metal selenides ($M_xSe_y$) have gained attention for their unique physical and chemical properties, especially those associated with the transition metal (M). Despite advancements in synthesis, fabricating these selenides is challenging due to their complex stoichiometry and high asymmetry. One such system is monoclinic iron selenide ($Fe_3Se_4$), which can be used in permanent-magnet technologies and serve as a model system for understanding magnetism. This study focuses on fabricating monoclinic $M_3Se_4$ (M = Fe, Co, or Ni) compounds via thermal decomposition, examining how solution chemistry influences their morphology and properties. With a Curie temperature of about 322 K, $Fe_3Se_4$ is ferrimagnetic, whereas $Co_3Se_4$ and $Ni_3Se_4$ are paramagnetic between 5 and 300 K. The latter two compounds also show higher catalytic activity for hydrogen evolution in water splitting, with maximum $H_2$-evolution rates of 1.01, 5.16, and 6.83 mmol $h^{-1}g^{-1}$ for $Fe_3Se_4$, $Co_3Se_4$, and $Ni_3Se_4$, respectively.

**Keywords:** $M_3Se_4$, transition metal selenides, thermal decomposition, magnetic and photocatalytic properties.

**Corresponding Author**

*E-mail: cs.gopinath@ncl.res.in, p.poddar@ncl.res.in.

# INTRODUCTION

Although transition metal chalcogenides (TMCs) are thought to be exciting materials with extensive phase diagrams, their complexity, lack of control over stoichiometry, and shape make them more difficult to manage than oxides.[1–5] The properties of these materials— thermal and electronic (electrical, magnetic and optical)—are highly delicate to their crystalline phase, stoichiometry, shape, and size.[6–11] The surface-to-volume ratio further improves the tunability of chemical and physical properties due to nano-dimensionality. Nanoscale phenomena provide an explanation for many of nature's wonders. A key challenge is controlling TMC phase formation chemistry. Transition metal selenides ($M_xSe_y$) are of special importance because of their electrical and thermal properties, which are determined by the transition metal (M), M-Se ratio, and crystallinity.[12–16] For applications in magnetic materials over some other materials, like magnetic read-heads, ferrofluids, high-density memory storage using magnetic media, or the growing significance of magnetic nanoparticles (NPs) in biomagnetic applications, precise control over the elemental ratio, size, shape, crystallinity, and assembly of NPs is essential because even small changes in these parameters have a comparatively greater impact on the applications.[17–24]

The Fe-Se system comprises four stable phases: orthorhombic $FeSe_2$, monoclinic $Fe_3Se_4$, hexagonal $Fe_7Se_8$, and tetragonal β-FeSe. In previous reports[7,25], we examined the intricacies of phase diagrams and stoichiometry in this system. Among four phases, $Fe_3Se_4$ exhibits unique ferrimagnetic properties due to ordered iron vacancies in alternating layers—in each layer, both the number and distribution of vacancies are different.[7] Ferrimagnetic ordering is thought to have its roots in the spins that are antiferromagnetically aligned in a plane next to ordered iron vacancies and ferromagnetically aligned inside each plane along the c-axis.[7,26,27]

Even in the absence of rare-earth or noble metal atoms, $Fe_3Se_4$ exhibits semi-hard magnetic characteristics below its Curie temperature at the nanoscale, making it a special material. Zhang et al.[27] found that when 90 kOe of the maximal external field was applied to nanoplatelets, the coercivity increased to around 40 kOe at 10 K from about 4 kOe at 300 K. Due to its monoclinic structure with ordered iron vacancies, $Fe_3Se_4$ exhibits a significant magnetocrystalline anisotropy ~1.0 x $10^7$ erg/cm$^3$ at 10 K.[26] At 300 K, the energy product ($BH_{max}$) for $Fe_3Se_4$ nanorods was ~4.38 kG Oe, which was improved by Mn-doping in an applied magnetic field of over 85 kOe to 10.22 kG Oe.[28] Additionally, our recent work[29] highlights how controlling the shape of $Fe_3Se_4$ NPs affects magnetic properties, with coercivity and remanence increasing with size. The $BH_{max}$ rises linearly with size, reaching 7.5 kG Oe for nanoplatelets and 7.1 kG Oe for nanorods at 300 K.

Due to the difficulty in fabricating highly asymmetric TMCs, the significant role of Fe in determining the magnetic characteristics of $Fe_3Se_4$ NPs has not yet been investigated. There have always been difficulties, curiosity, and possibilities related to the special and novel M-dependent characteristics that crystals exhibit. Therefore, to further comprehend the magnetic characteristics of $Fe_3Se_4$, a thorough study of monoclinic $M_3Se_4$ compounds (M= Fe, Co, and Ni) is intriguing and necessary. While $Co_3Se_4$ and $Ni_3Se_4$ are believed to be paramagnetic[30], this has not been experimentally verified. These compounds have also garnered attention for their high electrocatalytic activity in oxygen and hydrogen evolution reactions (OER and HER). [31–34] However, due to the difficulty in the synthesis, there are no or very few reports on the nanoscale fabrication of single phase $Co_3Se_4$. Furthermore, a binary photocatalyst $Ni_3Se_4$@CdS has been created, and when exposed to visible light, it produces hydrogen at a rate of 7.57 mmol $h^{-1}$ $g^{-1}$.[35]

$M_3Se_4$ (where M = Fe, Co, and Ni) is a transition metal selenide with a monoclinic crystal structure, $(M^{2+})_1(M^{3+})_2(Se^{2-})_4$. $M_3Se_4$ are crystallized in the monoclinic space group (SG)— I2/m ($Fe_3Se_4$ and $Ni_3Se_4$) and C2/m ($Co_3Se_4$) with distinct unit cell parameters (Figure 1). For $Fe_3Se_4$[36], the unit cell lattice parameters are a=6.20 Å, b=3.54 Å, and c=11.28 Å [α= γ= 90°, β= 91.807°] and for $Ni_3Se_4$[37] a=6.19 Å, b=3.63 Å, and c=10.45 Å [α= γ= 90°, β= 90.05°]. Whereas, for $Co_3Se_4$[38] the parameters are a=12.10 Å, b=3.57 Å, and c=6.18 Å [α= γ= 90°, β= 120.73°]. Each crystal structure has two geometrically different cation sites. For $Fe_3Se_4$— $[Fe^{3+}]_A$ $[Fe^{3+}Fe^{2+}]_B$ $(Se^{2-})_4$: A sites are tetrahedrally coordinated by Se atoms and are occupied exclusively by $Fe^{3+}$ cations, while B sites are octahedrally coordinated by Se atoms and are occupied by an equal number of $Fe^{2+}$ and $Fe^{3+}$ cations. For $(Co/Ni)_3Se_4$— $[(Co/Ni)^{2+}]_A$ $[((Co/Ni)^{3+})_2]_B$ $(Se^{2-})_4$: A sites are tetrahedrally coordinated by Se atoms and are occupied exclusively by $(Co/Ni)^{2+}$ cations, while B sites are octahedrally coordinated by Se atoms and are occupied exclusively by $(Co/Ni)^{3+}$ cations.

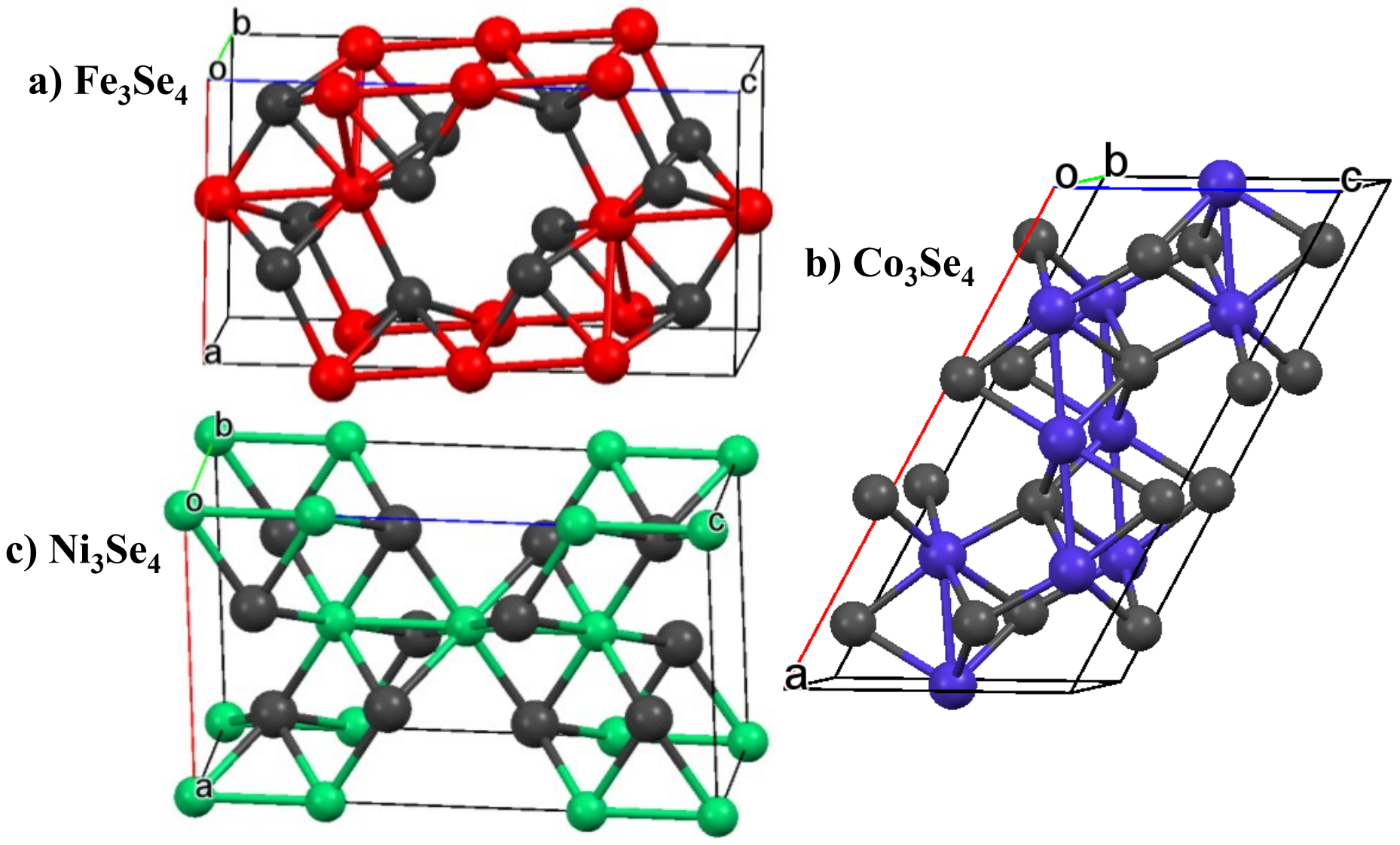

**Figure 1.** Schematic of the unit-cells of a) $Fe_3Se_4$, b) $Co_3Se_4$, and c) $Ni_3Se_4$ having monoclinic NiAs-type crystal structure. Red, blue, and green solid circles represent Fe, Co, and Ni cations. Dark grey solid circles represent Se anions.

This work utilizes thermal decomposition to achieve tunability of transition metals in complex $M_3Se_4$ compounds at the nanoscale. The thermal decomposition approach provides controlled synthesis, paving the way for regulated fabrication of other TMCs. Single-phase $M_3Se_4$ compounds have been synthesized and characterized using X-ray diffraction (XRD) and transmission electron microscopy (TEM). The study includes predictions of crystal habits using theoretical models and an exploration of the unique experimental morphologies. Thermogravimetric analysis (TGA) is used to assess decomposition behavior. Distinct magnetic and photocatalytic properties are observed with varying transition metals in monoclinic $M_3Se_4$ NPs, and the underlying reasons are discussed.

## EXPERIMENTAL SECTION

### Materials

Iron (III) acetylacetonate [$Fe(acac)_3$; 99.99%], cobalt (III) acetylacetonate [$Co(acac)_3$; 99.99%], nickel (II) acetylacetonate [$Ni(acac)_2$; 95%], selenium powder [Se; 99.99%], and oleylamine [OLA; 70%] were purchased from Sigma-Aldrich chemicals. All chemicals were used without further purification.

### Thermal Decomposition Based Synthesis of $M_3Se_4$ Compounds

A one-pot thermal decomposition approach was employed to create monoclinic $M_3Se_4$ compounds. Starting with $Fe_3Se_4$ NPs as a reference[29], $Co_3Se_4$ and $Ni_3Se_4$ NPs were synthesized under identical conditions, differing only in the transition metal precursor. In each reaction, 1.5 mmol of metal acetylacetonate [0.53 g for $Fe(acac)_3$ and $Co(acac)_3$; 0.38 g for $Ni(acac)_2$] and 2 mmol of Se powder [0.158 g] were used. These precursors were dissolved in 15 mL of oleylamine (OLA) in a 100 mL round-bottom flask under nitrogen at 30 °C. The temperature was increased from 30 °C to 120 °C, held for 30 minutes, then ramped to 200 °C at 2 °C/min and further to 330 °C at 5 °C/min, where it was maintained for 60 minutes. A thermometer was inserted inside the RB-flask for all reactions to ensure temperature stability within ± 2.0 °C. After cooling the reaction to room temperature (RT), the samples were rinsed with 20 mL of 2-propanol, centrifuged, and rewashed with 15 mL of n-hexane and 10 mL of 2-propanol. The resulting precipitate was dried in a vacuum at RT, afterwards utilized for further analysis.

### Synthesis of $M_3Se_4$/P25 composite

To investigate $H_2$ production activity, varying wt% of $M_3Se_4$ (M = Fe, Co, Ni) co-catalysts were loaded with titania (Degussa-P25) using the dry impregnation method.[39–40] The necessary concentrations of P25 and $M_3Se_4$ NPs were separately dissolved in ethanol and sonicated for 15 minutes. Afterward, the $M_3Se_4$ solution was incorporated to the P25 solution and sonicated for another 30 minutes to ensure homogeneous dispersion before drying for 12 hours at 60 °C. Figure S1 shows the final composite color for each $M_3Se_4$ weight percentage.

**Preparation of photocatalyst thin films over glass plates**

A 1.25 × 3.75 $cm^2$ glass plate (Figure S1b) was first washed with soap solution, then sonicated with de-ionized (DI) water and acetone for 45 minutes (15 minutes each), stored at 65 °C for three hours, and then utilized to create thin films. The photocatalyst thin films were prepared over the glass plates using the drop casting technique, which did not require the inclusion of binders.[39–40] To achieve a homogeneous dispersion of the catalyst, 1 mg of photocatalyst (P25 or $M_3Se_4$/P25) was incorporated to 1 mL of ethanol and sonicated for 30 minutes. A 100 µL micropipette was used to repeatedly drop-cast this catalyst dispersion onto the glass plate, and it was allowed to dry for 12 hours in the ambient conditions.

**Evolution of $H_2$ Via Photocatalysis**

In order to test the photo-catalytic hydrogen production capacity, one sun condition (100 $mW/cm^2$) was generated with a 300 W Xe Lamp and AM 1.5 filter (Newport solar simulator) source used at room temperature. In particular, a round-bottom glass flask with a volume of 72 ml was utilized as a reactor for the hydrogen evolution reaction. In the 72 ml reactor, which included 10 ml of methanol as sacrificial agent and 30 ml distilled water, the catalyst glass plate was placed exactly as it had been made. Then, a photo-catalyst-filled round bottom flask was purged with $N_2$ to exclude any dissolved oxygen in the solution and head space of reactor and subsequently exposed to one sun irradiation conditions. The generated $H_2$ was measured with a gas chromatograph (Agilent 7890). GC (Gas Chromatograph, Agilent 7890A) measurements verified that the only product in the gas phase was hydrogen.

## Characterization Techniques

Powder XRD measurements were conducted on a PANalytical X'PERT PRO instrument ($\lambda$ = 1.54 Å) over a $2\theta$ range of 10°-80°. The size, morphology, SAED, and lattice images were characterized using a FEI Tecnai T20 TEM operated at 200 keV. The powders were dispersed in n-hexane, drop-casted onto a carbon-coated copper TEM grid, and loaded into a single tilt sample holder. Mercury[41] 4.0.0 software was used to draw the internal crystal structures and packing diagrams of the $M_3Se_4$ compounds ($Fe_3Se_4$, $Co_3Se_4$, and $Ni_3Se_4$) crystal using the reported crystallographic information with COD numbers 1527086[35], 9012803[37], and 9009245[36], respectively. The indexed morphologies predicted using Bravais Friedel Donnay

Harker (BFDH), and Hartman Perdok (HP) methods were drawn using the *WinXMorph*[42] program. Thermal stability was assessed using a TGA SDT model Q600 of TA Instruments Inc. USA at a heating rate of 10 °C/min. Magnetic measurements were performed with a superconducting quantum interference device-based vibrating sample magnetometer (SQUID-VSM) manufactured by Quantum Design, Inc., San Diego, USA. The powder samples were precisely weighed and packed inside a plastic sample holder, which fits into a brass specimen holder provided by Quantum Design Inc. with a negligible contribution to the overall magnetic signal. The magnetization versus magnetic field (M–H) loops is collected at a rate of 50 Oe $s^{-1}$ in a field sweep from ± 60 kOe at the vibrating frequency of 40 Hz. The magnetization versus temperature (M−T) measurements was performed at a temperature sweep from 5 K to more than 300 K in a field of 100 Oe following standard field-cooled (FC) and zero-field cooled (ZFC) sequences. Catalyst materials and devices made were thoroughly characterized by various analytical methods. Diffuse reflectance UV-Vis measurement was carried by Shimadzu spectrophotometer (model-UV2550) with spectral grade $BaSO_4$ as a reference material. Hydrogen production was measured at zero potential in the wireless thin film configuration using one sun condition, as explained earlier.

# RESULTS AND DISCUSSION

### Structural and Morphological Investigations of $M_3Se_4$ compounds

Powder XRD was used to thoroughly examine the crystallinity and phase purity of the as-synthesised $M_3Se_4$ (where M=Fe, Co, and Ni) compounds. FullProf[43] software was then used to do the Reitveld refinement (Figure 2). The diffraction patterns showed no unidentified peaks, confirming the absence of secondary phases within the detection limits of the XRD technique. The patterns matched the reported standard diffraction data for $Fe_3Se_4$, $Co_3Se_4$, and $Ni_3Se_4$, with no ambiguous reflections. Table S1 presents the details of the refinement. $Fe_3Se_4$ XRD data is discussed in our previous work also.[29] The Scherrer [44,45] method was used to determine the crystallite sizes along various planes that corresponded to the strongest diffraction peaks: 42 ± 3 nm (-112) and 35 ± 3 nm (202) for $Fe_3Se_4$, 8 ± 2 nm (002) and 7 ± 2 nm (202) for $Co_3Se_4$, and 19 ± 2 nm (-112) and 17 ± 2 nm (-114) for $Ni_3Se_4$ along mentioned planes. All of the $M_3Se_4$ compounds with the various SG and the computed unit cell lattice parameters, as listed in Table S1, crystallized in the monoclinic crystal structure.

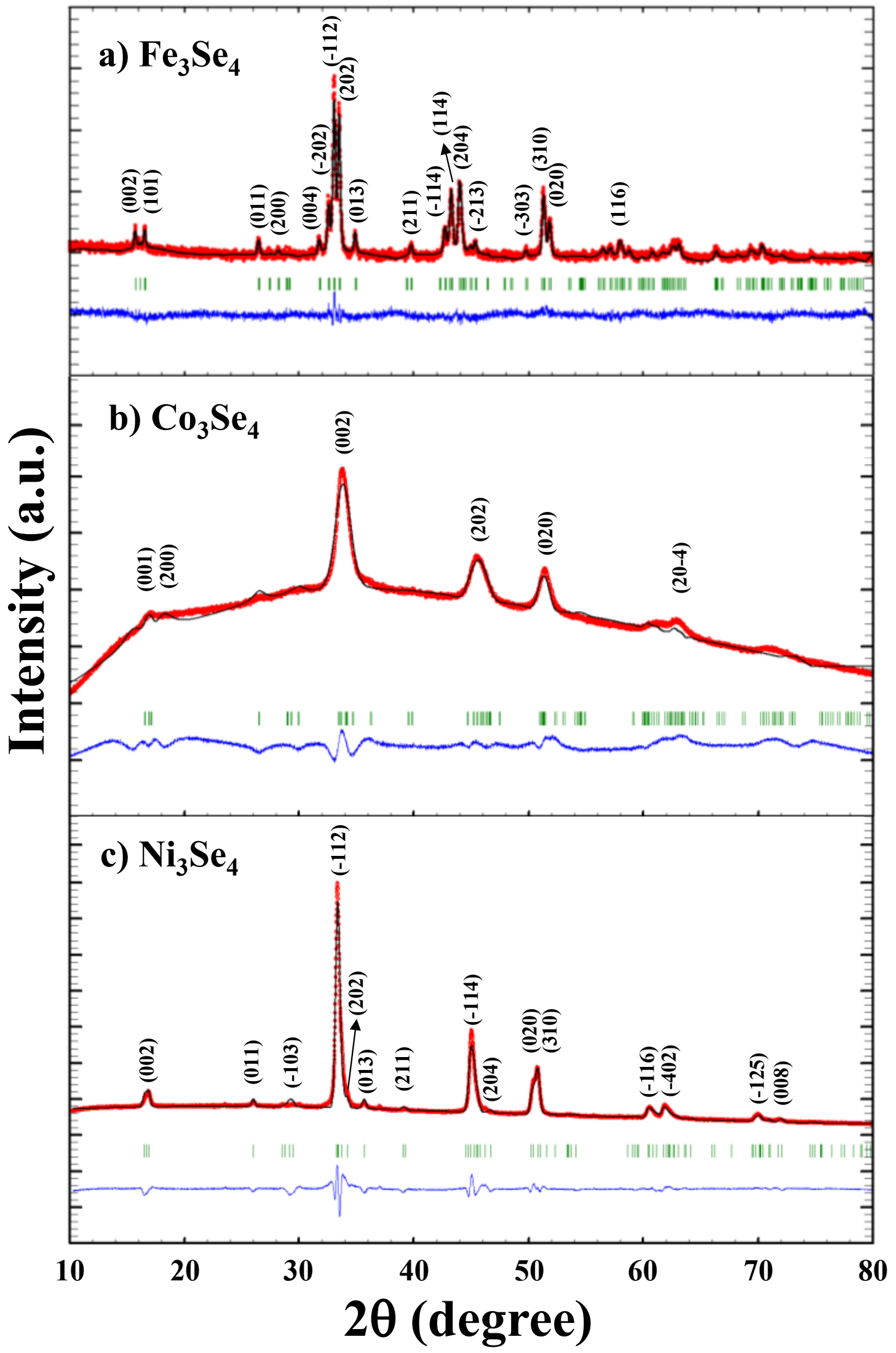

**Figure 2.** A comparison between experimentally obtained PXRD patterns (Observed Intensity $I_{obs}$ vs. 2θ) of a) $Fe_3Se_4$[29], b) $Co_3Se_4$, and c) $Ni_3Se_4$ (red lines) with calculated XRD curves (Calculated Intensity $I_{cal}$ vs. 2θ) obtained using Rietveld refinement method (black line), $I_{obs}$-$I_{cal}$, difference curve (blue line), and Bragg positions (green vertical lines).

Even under identical reaction conditions, the atomic arrangement in monoclinic $M_3Se_4$ compounds varies as M changes from Fe to Co, resulting in distinct SGs. Atoms are organized in the I2/m SG in $Fe_3Se_4$ and $Ni_3Se_4$, however in $Co_3Se_4$, they are placed in the C2/m SG. Despite being members of the same category, the atomic configurations of the two SGs differ, which has an impact on the electrical characteristics and crystal shape that result. Buerger (1951) classified phase transitions as reconstructive (slow) or displacive (rapid) based on structural changes.[46] Reconstructive transitions, like graphite to diamond[47], involve significant structural changes and require high activation energy. Displacive transitions involve minor bond distortions without breaking and occur readily with zero to minimal activation energy and cannot usually be prevented from occurring. Moreover, displacive transitions exhibit structural similarities. The two polymorphs have a symmetry relationship; the low-temperature polymorph has a low symmetry and is a subgroup of the high-temperature polymorph. Consequently, the displace transitions are frequently seen in the same class of SGs. For instance, the C-centered clinopyroxenes $LiScSi_2O_6$ and $ZnSiO_3$ exhibit a displace phase transition.[48] These transitions can only be detected using *in-situ* measurements. Displacive phase transitions are believed to occur in $Fe_3Se_4$ and $Ni_3Se_4$, but they are too rapid for ex-situ characterization. In $Co_3Se_4$, the reaction is incomplete, with atoms arranged in C2/m SG, and no phase transition occurs due to gel formation. Their energy and time for the phase shift are insufficient. Diffraction peak broadening suggests larger particle sizes in $Fe_3Se_4$ and $Ni_3Se_4$, while $Co_3Se_4$ particles remain small. Further microscopic analysis will provide additional insights into this phenomenon.

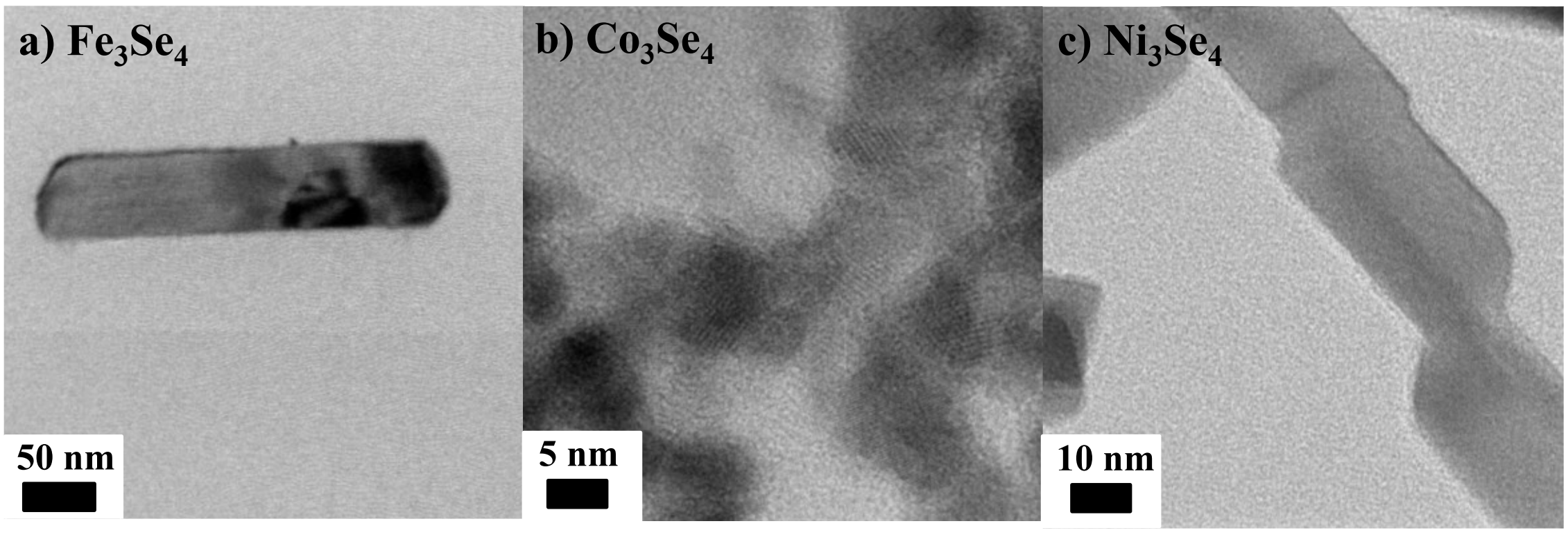

**Figure 3.** Typical TEM images of as-synthesized a) $Fe_3Se_4$, b) $Co_3Se_4$, and c) $Ni_3Se_4$ NPs (in the presence of OLA) show the rod-like features in $Fe_3Se_4$ and $Ni_3Se_4$. In contrast, small quasi-spherical-like or random-shaped features are observed in $Co_3Se_4$.

TEM analysis reveals that the as-synthesized $M_3Se_4$ compounds exhibit distinct morphologies: $Fe_3Se_4$ and $Ni_3Se_4$ form rod-like structures, while $Co_3Se_4$ appears as small quasi-spherical NPs (Figure 3). The NP shapes of $Fe_3Se_4$ and $Ni_3Se_4$ are similar, but $Co_3Se_4$'s morphology is markedly different. Scale bars indicate that all samples are within the nanometric range, with $Fe_3Se_4$ showing the fastest growth, producing large nanorods, followed by $Ni_3Se_4$ nanorods, and the smallest particles being the $Co_3Se_4$ quasi-spheres, despite identical reaction conditions.

All synthesized $M_3Se_4$ compounds underwent identical reactions, differing only by the metal acetylacetone precursor used. As previously reported [25,29], metal-OLA and selenium-OLA complexes are formed by metal and selenium precursors, respectively, and produce cations and anions. Compounds develop more quickly when there are more ions available. Selenium precursor reactivity will be similar to that covered in our earlier studies[25]— Se-OLA complexes provide $Se^{2-}$. The decomposition behavior of metal precursors is crucial, influencing the rate of cation and compound formation. Multiple metal acetylacetone's thermal breakdown was examined by Hoene et al.[49] The findings indicate that $Fe(acac)_3$, $Co(acac)_3$, and $Ni(acac)_2$ start to break down at temperatures over 184 ˚C, 213 ˚C, and around 200 ˚C, respectively. $Fe(acac)_3$ decomposes earliest, leading to the rapid growth of large $Fe_3Se_4$ nanorods. In contrast, $Co(acac)_3$ decomposes at higher temperatures, resulting in smaller $Co_3Se_4$ nanoparticles. $Ni(acac)_2$ falls in between, producing intermediate-sized $Ni_3Se_4$ nanorods. The metal precursor thus plays a key role in determining NP size in $M_3Se_4$ compounds.

After studying $M_3Se_4$ sizes, a detailed examination of the significant differences in $M_3Se_4$ shapes, as seen in TEM images, reveals that $Fe_3Se_4$ and $Ni_3Se_4$ NPs exhibit rod-like shapes, while $Co_3Se_4$ forms small quasi-spherical shapes under the same experimental conditions.

The theoretical analysis offers insights into these differences. Two primary approaches, Bravais Friedel Donnay Harker[50–52] (BFDH) and Hartman Perdok[53–55] (HP), are typically used to predict crystal morphology. The specifics of the crystal habit calculations are covered in supporting information (Figure S2-S7 and Table S2-S3). The BFDH model, based on crystal parameters, shows that $Co_3Se_4$ has different crystal parameters and space groups (SG) compared to $Fe_3Se_4$ and $Ni_3Se_4$, leading to a distinct crystal habit. In contrast, the HP model, which relies on bond energies, predicts minimal differences in morphology due to the similar bond energies across the compounds, with only slight directional variations. The variation in crystal parameters primarily influences the observed differences in NP shapes.

Materials behave more complexly in real-life reactions than they do in theoretically simplified models considering the effect of outside variables. Figure S8 illustrates the morphology of $Fe_3Se_4$ in both experimental and theoretical circumstances. According to both theoretical models, $Fe_3Se_4$ has a rod-like crystal habit, with the longitudinal sides being formed by the (001), (101), and (-101) planes. This prediction aligns with the observed experimental $Fe_3Se_4$ nanorods, which display lattice fringes at 5.6 Å and 2.6 Å corresponding to (002) and (202) planes on the longitudinal sides, confirming the rod-like shape. Our earlier study also discusses $Fe_3Se_4$ TEM observations in detail.[7,25,29] In conclusion, the theoretical and experimental morphologies are consistent under the given reaction conditions.

In contrast, the crystal habit of $Co_3Se_4$ predicted by the BFDH and HP models differs significantly— Figure S9 illustrates the BFDH's rectangular-box shape and the HP model's compact cylindrical shape. Experimentally, $Co_3Se_4$ NPs are observed as small quasi-spherical shapes with (200) and (400) planes with lattice fringes at 5.2 Å and 2.6 Å. The $Co_3Se_4$ has only begun its formation process under our reaction conditions, resulting in very tiny NPs in comparison to other $M_3Se_4$. When the reaction has started, the crystallographic model will dominate the crystal shape at low energy. Later, depending on the reaction environment, bond energy theories will also play a significant role. Since the number of intermolecular interactions in a crystal increases with its face area.[56] As a result, the growth rate increases in that direction, which in turn leads to a decrease in MI, and vice versa. As the (100) planes in $Co_3Se_4$ have a smaller facial area, interaction and growth in this face occur more slowly, leading to a greater MI of (100). Thus, at the initial stage of reactions, the crystallographic morphology dominates; consequently, with (200) and (400) planes on the upper side, the observed $Co_3Se_4$ NPs have a quasi-spherical form and are close to the BFDH model.

The morphology of the $Ni_3Se_4$ compound closely resembles that of $Fe_3Se_4$, both theoretically and experimentally. $Ni_3Se_4$ shares the rod-like characteristics of (001), (101), and (-101) on the longitudinal sides according to both theoretical models, as was previously mentioned. Furthermore, as seen in Figure S10, the synthesized $Ni_3Se_4$ NPs are likewise shaped like a rod, with a 2.6 Å lattice fringe space in the (202) plane on the longitudinal side, which is fairly comparable to that of $Fe_3Se_4$. In summary, since both theoretical models predict a rod-like morphology for $Ni_3Se_4$, the experimental results align closely with these predictions, just as they do for $Fe_3Se_4$.

Therefore, after taking into account the crystal habits of $M_3Se_4$ as well as the thermal decomposition of metal acetylacetonate, it is appropriate to conclude that, in the case of $Fe_3Se_4$, the longer reaction time resulted in a largest-size rod shape NPs, whereas in the instance of $Co_3Se_4$, the reaction barely began and produced small particles with a crystallographic crystal habit. Between these two extreme circumstances, $Ni_3Se_4$ has a rod-like shape but is smaller in size than $Fe_3Se_4$.

**Thermal Study**

To investigate the decomposition behaviour of the as-synthesised NPs, the TGA of $M_3Se_4$ NPs was performed under a flowing nitrogen environment by heating the specimens over 25 °C to 1000 °C at a rate of 10 °C/min. Every sample exhibited an independent weight-loss tendency, illustrated in Figure S11. For $Fe_3Se_4$, $Co_3Se_4$, and $Ni_3Se_4$, the weight loss was 3%, 15%, and 5%, respectively, with the initial step (<460 °C) being associated with the loss of organic particles (evaporation of capped OLA). The second step was devoted to the breakdown of $M_3Se_4$, which begins at temperatures over 700 °C (weight loss ~4%, 20%, and 8%) and progressively drops to temperatures up to 1000 °C, with perceived (cumulative) weight loss of ~20%, 43%, and 32%, respectively. Here, the $M_3Se_4$ compounds decomposed into another $M_{1-x}Se$-based molecule, which was followed by further decomposition. $Fe_3Se_4$ TGA data is covered in our previous work also[25].

**Significant Impact of M on Magnetic Properties at Nanoscales in Monoclinic $M_3Se_4$ Compounds**

Using SQUID-VSM, the impact of transition metals (M) on the nanoscale magnetic characteristics of $M_3Se_4$ compounds was evaluated. The $Fe_3Se_4$ magnetic properties are taken from our previous work[29]. Figure 4 presents the temperature-dependent magnetization (M-T) curves for $M_3Se_4$ NPs in zero-field cooling (ZFC) and field cooling (FC) modes with an applied magnetic field of 100 Oe. For $Fe_3Se_4$ (Figure 4a), the M-T curve reveals a bifurcation below 330 K and a Curie transition temperature of ~322 K, indicating a ferrimagnetic phase below this temperature. $Co_3Se_4$ and $Ni_3Se_4$ (Figures 4b and 4c) show overlapping ZFC and FC curves across the entire temperature range (5K to more than 300 K), suggesting paramagnetic behavior with magnetic transition temperatures below the measurement range. Figure S12 further demonstrates that the magnetic susceptibility of both $Co_3Se_4$ and $Ni_3Se_4$ decreases with increasing temperature. The inverse of the magnetic susceptibility plot reveals that the extrapolation to zero temperature fails to obey the Curie Law. Instead, both the samples follow the Curie-Weiss Law as intercepts are above 0 K.[57]

The magnetic field-dependent magnetization (M-H) of $M_3Se_4$ NPs was analyzed at 300 K and 10 K to explore their magnetic behavior. As previously reported[27,29], Figure 5a displays the hysteresis loops with typical non-linear behavior indicative of the ferrimagnetic nature of $Fe_3Se_4$ NPs at 300 K. In line with earlier research[26–28], the magnetization remains unsaturated even at the maximum applied field of 60 kOe where the extremely high anisotropy field or the system's noncollinear spins prevented saturation even at 90 kOe. At 300 K, the coercivity ($H_C$) and remanence magnetization ($M_R$) values are 2.3 kOe and 2 emu/g, respectively, as shown in the inset of Figure 5a. At 10 K, the M-H curve remains unsaturated at 60 kOe (Figure 5a'), with coercivity values increasing significantly to 35.5 kOe—nearly 30 times greater than at 300 K—because of things like ferrimagnetism, an increase in total effective magnetic anisotropy, and a decrease in the thermal activation energy that gives up more and more spins to orient in the field direction.[29] Additionally, at 10 K, the $M_R$ increases from 2 emu/g at 300 K to 7.6 emu/g.

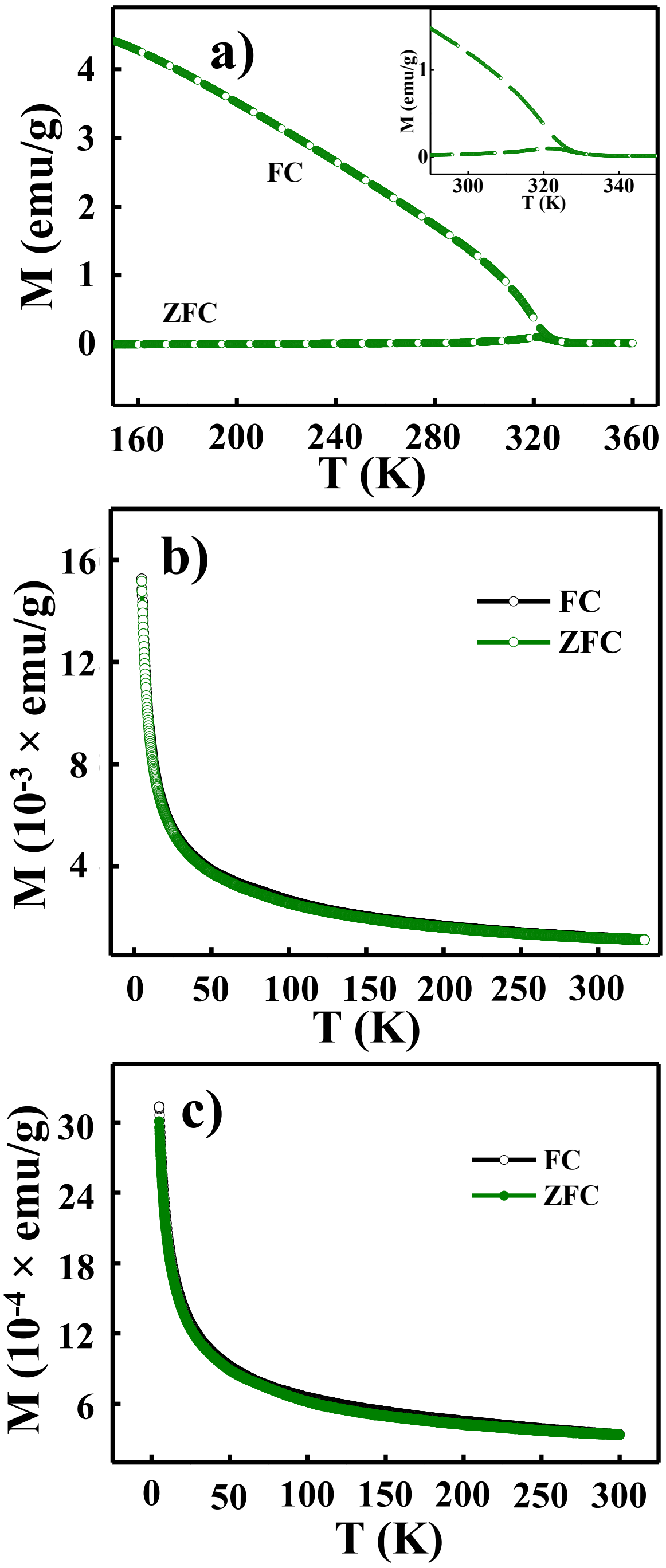

**Figure 4.** Comparison of magnetization versus temperature curves (in ZFC and FC modes) for all the as-synthesized $M_3Se_4$ NPs. a) to c) curves correspond to the samples— $Fe_3Se_4$[29], $Co_3Se_4$, and $Ni_3Se_4$, respectively, in the applied magnetic field of 100 Oe.

In contrast to $Fe_3Se_4$, both $Co_3Se_4$ and $Ni_3Se_4$ exhibit linear M-H curves at 300 K, indicating a paramagnetic nature, which persists at 10 K as shown in Figures 5b-c and 5b'-c'. As a result, these compounds exhibit paramagnetic behavior at both observed temperatures (300 K and 10 K), with the magnetic transition occurring below 10 K. The magnetic properties of $M_3Se_4$ compounds with a monoclinic crystal structure vary significantly with the transition metal: $Fe_3Se_4$ is ferrimagnetic at both 300 K and 10 K, while $Co_3Se_4$ and $Ni_3Se_4$ remain paramagnetic across the same temperature range. While the magnetic transition temperature in $Co_3Se_4$ and $Ni_3Se_4$ is lower than the lowest measured temperature (5 K), the Curie transition temperature of $Fe_3Se_4$ is close to 322 K. The magnetic characteristics change significantly even if the transition metal is changed from Fe to Co/Ni in the same structure. The explanation for this intriguing discovery resides in the magnetic characteristics of $M_3X_4$ (where M = Fe, Co or Ni, and X = O, S or Se) compounds.

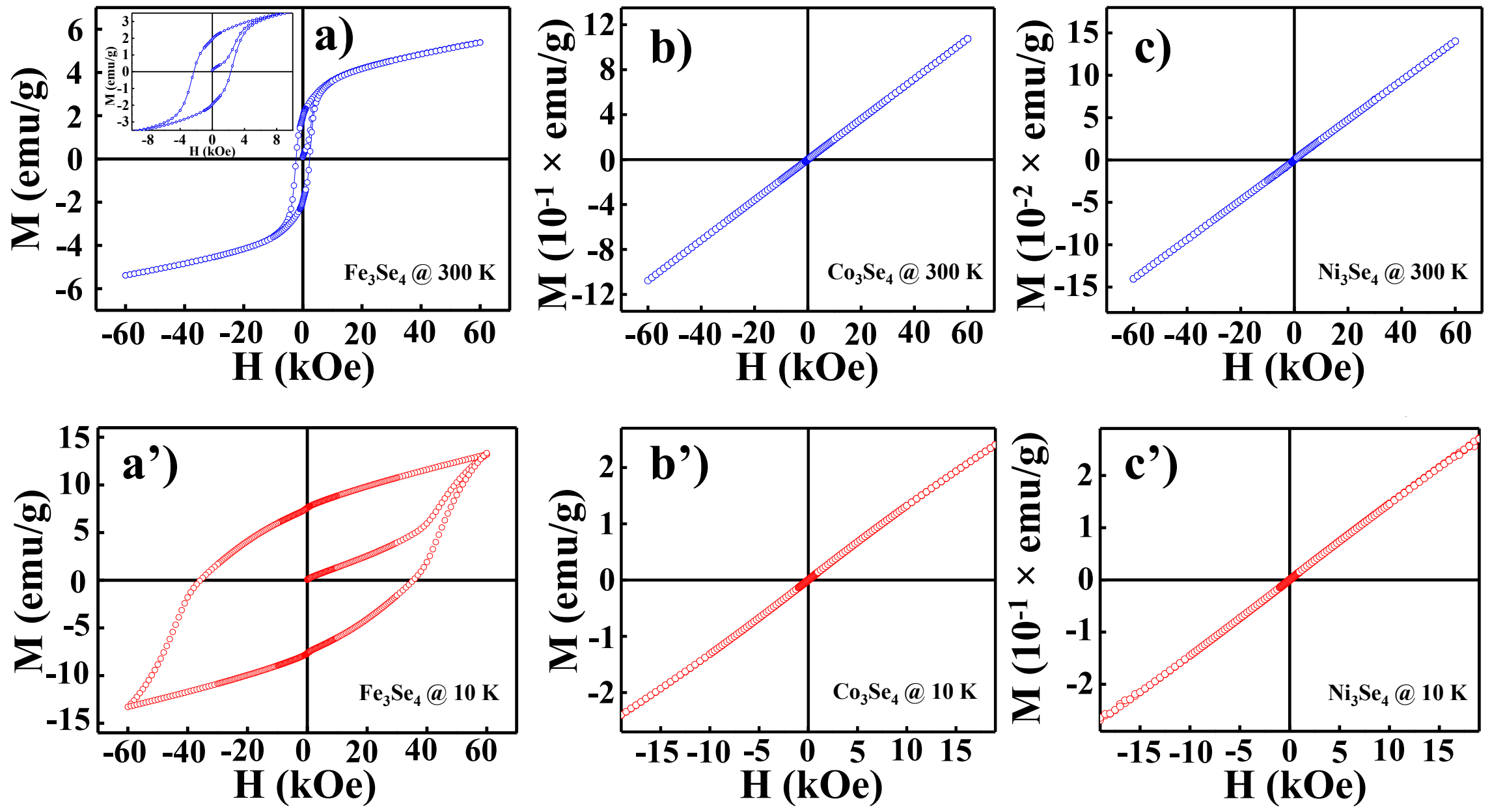

**Figure 5.** Magnetization vs. external magnetic field hysteresis loops at 300 K (blue curve) and 10 K (red curve) of $M_3Se_4$ NPs. a) to c) blue loops correspond to the samples— $Fe_3Se_4$[29], $Co_3Se_4$, and $Ni_3Se_4$, respectively at 300 K. a') to c') red loops correspond to the samples— $Fe_3Se_4$[29], $Co_3Se_4$, and $Ni_3Se_4$, respectively at 10 K.

The magnetic transition temperature (Curie or Neel temperature) of $M_3X_4$ compounds is displayed in Table S4.[58–81] The key question is what will happen to magnetic characteristics if M (Fe, Co, or Ni) remains constant while X (O, S, or Se) changes, or vice versa.

1) $Fe_3X_4$ is used as a case study in order to comprehend the first point. The anion's atomic size grows as X shifts from O to S to Se, raising the lattice parameters and lowering the exchange interaction.[82] The covalency effect and delocalization of 3D electrons also contribute to the reduction of the magnetic moment.[83] Overall, as X's atomic size increases, magnetization decreases. As a result, when X's atomic size increases, the thermal energy required to randomly orient the magnetic spins reduces, and thus magnetic transition temperature reduces from O to Se. The Curie temperature reduces from ~858 K for $Fe_3O_4$ [58–62] to ~314 K for $Fe_3Se_4$.[27,80]

2) In the second point, maintaining X constant and changing the M— $M_3O_4$ is used as a case study. The structure of $Fe_3O_4$ is inverted spinel. $Fe^{2+}$ is diamagnetic as it is octahedral, having no unpaired electrons ($t_{2g}^6$). The arrangement of atoms in $Fe_3O_4$ compounds results in ferrimagnetism because $Fe^{3+}$ is both tetrahedral ($e^2$ $t_2^3$) and octahedral ($t_{2g}^3$ $e_g^2$), with five unpaired electrons each, giving rise to a high magnetic moment.[84,85] The normal spinel structure is occupied by the other two components, $Co_3O_4$ and $Ni_3O_4$.[86–89] The $Co^{3+}$ in $Co_3O_4[Co^{2+}(Co^{3+})_2O_4]$ is in the octahedral state and has no unpaired electrons (low spin: $t_{2g}^6$), it possesses no magnetic moments. While, the $Co^{2+}$ is in a tetrahedral configuration (high spin: $e^4$ $t_2^3$) and has a 3.26 $\mu_B$ magnetic moment at 4.2 K. As a result, only $Co^{2+}$ participates to the magnetism of the $Co_3O_4$, which results in antiferromagnetic (AFM) behavior with a low Neel temperature (~ 40 K). As the overall magnetic moment is very small and only a small amount of thermal energy is required to randomly orient the magnetic moments.[63] In $Ni_3O_4$, $Ni^{2+}$ is in a tetrahedral state (high spin: $e^4$ $t_2^4$), and $Ni^{3+}$ is in an octahedral state (low spin: $t_{2g}^6$ $e_g^1$). As both cations have unpaired electrons and are positioned so that each layer's total magnetization varies in magnitude, results in a ferrimagnetic nature with a Curie temperature of close to 808 K. [64–68] The existence of more unpaired electrons in two types of Ni ions causes $Ni_3O_4$ to have more resultant magnetization than $Co_3O_4$, which means that the magnetic transition temperature will be much greater in $Ni_3O_4$ compared to $Co_3O_4$. As each $Fe^{3+}$ has five unpaired electrons, the resulting magnetization is quite strong in the case of $Fe_3O_4$, which also results in a high Curie temperature (~858 K). [58–62]

Combining these insights, we can address the behavior of $M_3Se_4$ compounds. Since the magnitude of the octahedral crystal field stabilizing energy reduces with increasing donor atom size. Thus, the Co and Ni ions move towards high spin states as the compounds transition from $(Co/Ni)_3O_4$ to $(Co/Ni)_3Se_4$. The exchange interaction declines from O → S → Se, resulting in a drop in total magnetization and magnetic transition temperature (as seen in $Fe_3X_4$—the magnetic transition temperature reduces from O to Se). Some research on Co-based compounds has shown that $Co_3S_4$ is a temperature-independent paramagnetic[75], while other studies assert that it possesses an antiferromagnetic with a Neel temperature of about 58 K.[76,77] Consequently, it is appropriate to draw the conclusion that magnetization has either reduced or remained relatively constant from O to S in Co-based compounds. Now when we move from S to Se, the exchange interaction decreases even more, lowering the magnetization in the instance of $Co_3Se_4$ to the point where the magnetic transition temperature falls below the detection limits. In the case of Ni-based compounds, it significantly drops from 808 K[64–68] ($Ni_3O_4$) to about 20 K[78,79] for $Ni_3S_4$ and then falls below the detection limit for $Ni_3Se_4$. The magnetic transition temperature in $Co_3Se_4$ and $Ni_3Se_4$ is therefore below the lower limit of the measurement temperature.

**M's Effect on Photocatalytic Properties as a Cocatalyst in Monoclinic $M_3Se_4$ Compounds**

The XRD patterns of each catalyst are shown in Figure S13, which further confirms the catalysts' phase structure as manufactured. It displays clear diffraction peaks, proving that $TiO_2$ and $M_3Se_4$ were effectively synthesized and that, within the bounds of laboratory XRD, they have acceptable crystallinity. In case of Fe/$Ni_3Se_4$ cocatalyst, XRD reveals small shoulder and peaks above 5 wt% of corresponding cocatalyst.

The UV-vis spectroscopy and Tauc plot were used to investigate the samples' optical characteristics and band gap structure.[90] Figure S14 a-c shows that both pure $TiO_2$ and $M_3Se_4/TiO_2$ samples exhibit a significant absorption in the UV light region, which is characteristic of $TiO_2$. The as-prepared $M_3Se_4$ materials display a strong absorption in the entire visible light region, which corresponds to their narrow band gap. Green curve observed for 5% $Ni_3Se_4$ integrated $TiO_2$ suggests the visible light absorption from orange-red region of visible light spectrum and below. Similarly, blue curve associated with $Co_3Se_4$ also indicating the light absorption for almost the entire visible range. While red curve associated with $Fe_3Se_4$ hinting the light absorption to extend into NIR with possibly smaller band gap than the other two counterparts. Indeed, all the three chalcogenides are known to be a low band-gap semiconductors and visible light absorption is easily facilitated. In view of visible light absorption, they also act as visible light absorption component after integration with $TiO_2$, apart from the role of co-catalyst. Nonetheless, large amount of co-catalyst loading on $TiO_2$ itself would prevent efficient light absorption, which is counterproductive to light to charge carrier generation applications, such as photocatalysis. The samples' band gap energy was assessed using the Tauc plots [90], which were produced using the formulas of

$$[F(R).h\nu]^n = A\ (h\nu - Eg) \text{ and } [\alpha h\nu]^n = A\ (h\nu - Eg)$$

(A is a constant, n = 2 for the direct band gap semiconductor).

The linear extrapolation of Tauc plots yielded a band gap energy of approximately 3.24 eV for $TiO_2$ (Figure S14a'-c'). However, 5 wt % chalcogenide integrated $TiO_2$ shows a comparable band-gap between 3.1 and 3.2 eV, indicating that they are comparable.

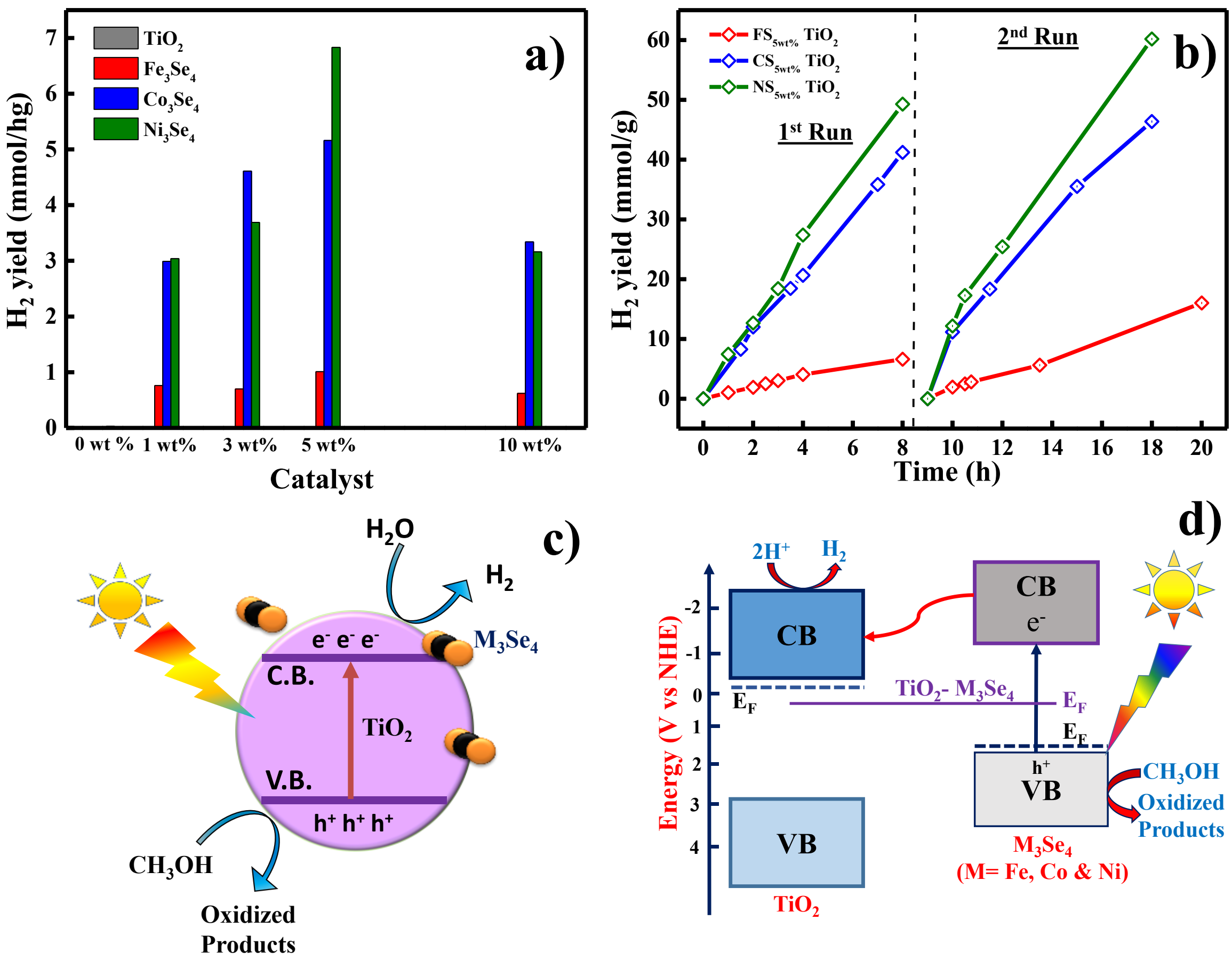

**Figure 6**. a) Photocatalytic solar hydrogen production with $TiO_2$ and $M_3Se_4$ as a co-catalyst for 4 h with different wt% of cocatalyst. b) Time dependent and cumulative photocatalytic hydrogen evolution of $M_3Se_{4\ (5wt\%)}$-$TiO_2$ catalysts. Cycling test of $M_3Se_{4\ (5wt\%)}$-$TiO_2$ for the photocatalytic $H_2$ evolution under one sun irradiation conditions. After 1$^{st}$ run of 8 h, the reactor was evacuated to remove the $H_2$ generated before the commencement of irradiation for 2$^{nd}$ run. c-d) The proposed photocatalytic mechanism for hydrogen evolution water splitting under visible light irradiation.

Using methanol as the sacrificial reagent, the photo-catalytic activity of $TiO_2$ and $M_3Se_4$ $_{(1\ to\ 10\ wt\%)}$-$TiO_2$ samples was examined under visible light irradiation. As shown in Figure 6a, pure $TiO_2$ exhibited a very low photocatalytic activity with a meagre solar hydrogen production yield of 0.029 mmol $h^{-1}g^{-1}$ under one sun conditions. Following the combination of $TiO_2$ and $M_3Se_4$, the composite catalyst demonstrates more photocatalytic activity than with $TiO_2$ alone. This is due to two reasons, namely, significant visible light absorption by co-catalyst which enhances the generation of electron-hole pairs. Second the heterojunctions created between $TiO_2$ and $M_3Se_4$ helps to rapidly separate the electron-hole pair generated, and the electrons are transferred to the CB of $TiO_2$. Additionally, this significantly reduces the ability of photo-produced charge carriers to recombine. From $M_3Se_{4(1\ wt\%)}$-$TiO_2$ to $M_3Se_{4(5\ wt\%)}$-$TiO_2$, the $H_2$ generation activity progressively rises and then falls for all three chalcogenides. Therefore, the optimal loading quantity of $M_3Se_4$ is 5 wt % and the highest $H_2$-evolution values was 1.01, 5.16 and 6.83 mmol $h^{-1}g^{-1}$ for $Fe_3Se_4$, $Co_3Se_4$ and $Ni_3Se_4$, respectively. It illustrates how the amount of $M_3Se_4$ that is loaded onto the surface of $TiO_2$ NPs is essential for increasing the samples' photo-catalytic water-splitting activity because too much $M_3Se_4$ may hinder the swift transfer of electrons.

It is appropriate to conclude that the effect of M in $M_3Se_4$ composition of co-catalysts is strong on the photocatalytic activity for $H_2$ evolution. The activity of bare $TiO_2$ NPs for photocatalytic $H_2$ evolution was found to be fairly low (Figure 6a). After adding $M_3Se_4$ as a cocatalyst the activity increased prominently. In case of $Fe_3Se_4$ as a cocatalyst, the highest achieved activity is 1.01 mmol $h^{-1}g^{-1}$, while in case of other two cocatalysts, it is increasing 5 to 6 times greater than that of $Fe_3Se_4$ and nearly 180 to 235 times better than that of $TiO_2$ NPs. This manifests that the $Ni_3Se_{4\ (5\ wt\%)}$-$TiO_2$ and $Co_3Se_{4\ (5\ wt\%)}$-$TiO_2$ materials prepared by above mentioned process has a close contact between $TiO_2$ and co-catalyst. TEM results shown earlier indicate the possibility of integration of $TiO_2$ with $Fe_3Se_4$ is low, due to its large particle size than the other two chalcogenides; however, $Co_3Se_4$ and $Ni_3Se_4$ shows comparable particle size range. Additionally, the advantage of nanorod with the $Ni_3Se_4$ cannot be ignored, as it can interact with several $TiO_2$ particles. A careful analysis of results shown in Figure 6a suggests the hydrogen production per unit weight of co-catalyst is the highest with $Ni_3Se_4$ at 5 wt %, while the same is true with $Co_3Se_4$ at 5 wt %. Hence it may be concluded that both the co-catalysts are comparable in terms of activity, however due to rod-shape of $Ni_3Se_4$, it shows slightly higher activity.

As demonstrated in Figure 6b, the photo-catalytic stability of the water-splitting process of $M_3Se_{4\ (5\ wt\%)}$-$TiO_2$ samples was evaluated. Two cycles of the stability test were carried out. Figure 6b makes it evident that the $M_3Se_{4\ (5\ wt\%)}$-$TiO_2$ catalyst sustains a high rate of hydrogen production per cycle, indicating that it has good stability. The primary reason for this is that the addition of $M_3Se_4$ as an electron promoter could continuously speed up the separation and transfer of photo-generated charge carriers on the surface of $TiO_2$.

The photo-catalytic reaction process and electron transfer mechanism under visible light irradiation were hypothesized based on the aforementioned results, presented in Figure 6c-d. Irradiation with one sun conditions produces photo-induced electron-hole pairs on both $TiO_2$. The rate of $H_2$ evolution for the $TiO_2$ without any co-catalyst was very low due to the fast recombination of photo-excited electron-hole pairs, despite the fact that the conduction band (CB) potential of $TiO_2$ is more negative than the reduction potential of $H^+/H_2$. When exposed to visible light, the electrons in the $M_3Se_4$ valence band (VB) became excited and swiftly moved to the CB position of $TiO_2$. To effectively prevent photo-generated electron and hole recombination, $M_3Se_4$ particles were distributed on the surface of $TiO_2$ as a co-catalyst to speed up charge carrier separation and transfer; in other words, charge carrier recombination would be effectively prevented. Therefore, adding $M_3Se_4$ can improve the HER's active site and speed up electron transfer, increasing the composite catalyst's capacity to produce hydrogen. A large jump in solar hydrogen production after introduction of $M_3Se_4$ on $TiO_2$ underscores the integrated nature of the final photocatalyst composite. 200 times increase in activity with $Ni_3Se_{4\,(5\,wt\%)}$-$TiO_2$ than $TiO_2$ underscores an efficient separation of charge carriers at the heterojunctions between the components in the former. This integration aspect necessarily shifts the Fermi level of the composite, which is different from the constituent components, and shown schematically in Figure 6d. This makes the electron transfer very facile and ultimately enhances the solar hydrogen production.

Further, these compounds are very promising in various scientific domains. Researchers are focusing more on these areas because $Ni_3Se_4$ is also a viable catalyst and $Co_3Se_4$ recently attracted a lot of attention due to its high OER values through electrocatalysis and photoelectrochemical water splitting.[31–35] The other characteristics of these pure as-synthesised chemicals will therefore be fascinating and crucial to investigate further in the future.

## CONCLUSION

Transition metal selenides are difficult to fabricate, and it is more challenging to create compounds with more asymmetry. The thermal decomposition approach was used in this work to manufacture the $M_3Se_4$ compounds (where M = Fe, Co, or Ni) in a monoclinic crystal structure with unique SG, resulting in a variety of morphologies. The atoms of $Co_3Se_4$ are arranged in C2/m, whereas the SG of $Fe_3Se_4$ and $Ni_3Se_4$ is I2/m. It is most likely the displacive transition that causes the unique SG. Theoretical models (BFDH and HP) assist in the understanding of the observed varying morphologies. Additionally, the impact of transition metal (M) on the magnetic characteristics of $M_3Se_4$ NPs is also examined. The Curie temperature of the ferrimagnetic $Fe_3Se_4$ is close to 322 K. $Co_3Se_4$ and $Ni_3Se_4$, on the other hand, are temperature-independent paramagnetic throughout the entire measurement temperature range (5 K to 300 K). In the latter two compounds, the magnetic transition temperature is so low that it cannot be detected within measurement limitations. However, these compounds as a cocatalyst shows prominent activity. The optimal loading quantity of $M_3Se_4$ is 5 wt % and the maximum $H_2$-evolution values were 1.01 mmol $h^{-1}g^{-1}$, 5.16 mmol $h^{-1}g^{-1}$ and 6.83 mmol $h^{-1}g^{-1}$ for $Fe_3Se_4$, $Co_3Se_4$ and $Ni_3Se_4$, respectively. The activity is 5 to 6 times higher than that of $Fe_3Se_4$ in other two compounds and almost 180 to 235 times better than that of $TiO_2$ NPs. Furthermore, the catalyst $M_3Se_{4\ (5\ wt\%)}$-$TiO_2$ sustains a high rate of hydrogen generation per cycle, indicating that it has good stability. To sum up, the work offers a suitable technique for creating $M_3Se_4$ NPs with unique M that have a range of magnetic and catalytic characteristics. This approach can be expanded to include the majority of TMCs.

## ASSOCIATED CONTENT

### Supporting Information

Rietveld refinement, BFDH, HP, packing diagram of $M_3Se_4$, magnetic properties, XRD and UV-vis of catalyst.

### Conflict of Interest Statement

The author (M.G.) declares that there is no conflict of interest regarding the publication of this work. This research is part of the author's Ph.D. thesis, which has been submitted and is publicly available on the institute's website. The reference for the thesis is https://dspace.ncl.res.in/xmlui/handle/20.500.12252/6067; titled as "Physics of transition metal chalcogenides based on deeper investigation of their phase-diagram and crystal growth mechanism".

## AUTHOR INFORMATION

### Corresponding Author

*E-mail: cs.gopinath@ncl.res.in, p.poddar@ncl.res.in.

### Orchid

Pankaj Poddar: 0000-0002-2273-588X

### Notes

The authors declare no competing financial interest.

## ACKNOWLEDGMENTS

One of the authors—MG acknowledges the support from the University Grants Commission (UGC), India, for providing the Senior Research Fellowship (SRF).

**SUPPORTING INFORMATION**

# Transition Metal-Driven Variations in Structure, Magnetism, and Photocatalysis of Monoclinic $M_3Se_4$ (M = Fe, Co, Ni) Nanoparticles

Monika Ghalawat[†,‡], Inderjeet Chauhan[†,‡], Dinesh Singh[†,‡], Chinnakonda S. Gopinath[†,‡]*, Pankaj Poddar[†,‡]*

[†] *Physical & Materials Chemistry Division, CSIR-National Chemical Laboratory, Pune 411008, India*

[‡] *Academy of Scientific and Innovative Research (AcSIR), Sector 19, Kamla Nehru Nagar, Ghaziabad, Uttar Pradesh- 201 002, India*

Number of figures: **14**

Number of schemes: **00**

Number of tables: **04**

**Crystal Habit of $M_3Se_4$ Compounds using BFDH and HP Model**

Usually, two successful approaches—Bravais Friedel Donnay Harker[1–3] (BFDH) and Hartman Perdok[4–6] (HP) have been used to predict the morphology of crystals. These models have been efficiently used to predict the crystal morphology of numerous organic and inorganic crystals such as ZnO[7], FeOOH[8], Nd: $LaVO_4$[9], $BaSO_4$[4–6], $GdVO_4$[10], $GeO_2$[11], $CaCO_3$[12], triglycine sulfate[13], benzophenone[14], $Pb_{17}O_8Cl_{18}$[15], etc. and helped in understanding the experimental morphologies. Here too, we looked at the crystal habit of $M_3Se_4$ compounds by using these models to explain experimental observations.

To predict the morphology of a crystal— A. Bravais[1] (1866), G. Friedel[2] (1907), and J.D.H. Donnay and D. Harker[3] (1937) proposed the theories using crystal lattice geometry and combinedly named as BFDH model. Here, the d-spacing (interplanar spacing) and SG symmetry operations are used to simulate the crystal habit for a particular crystal. Thus, as per the BFDH model, the faces with larger d spacing and higher density will grow slower and are most energetically stable. The unit cell dimensions and SG of the related crystal structure are used to calculate the relative growth rate (RGR) and morphological significance (MI) of various planes that decide the crystal's final form. The RGR and MI of low-index planes of $M_3Se_4$ compounds are explained in Table S2 using the BFDH model, and the crystal habit of the corresponding compounds was estimated using the WinXmorph[16] software utilizing these RGR and MI.

For $Fe_3Se_4$ crystal, (001) and (00-1) faces are of highest MI, followed by (-101), (10-1), (101), and (-10-1) planes. The resultant crystal habit of $Fe_3Se_4$ is a rod-like shape with longitudinal side— (001) and (00-1) faces strongly visible, followed by (-101), (10-1), (101), and (-10-1) planes. The other planes ((011), (0-11), (01-1), (0-1-1), (110), (-110), (1-10), (-1-10)) are on the edges of the rod as shown in Figure S2a.

Further, the $Co_3Se_4$ crystallographic crystal habit is predicted. This compound is unlike $Fe_3Se_4$ because it has a distinct unit cell parameter and SG; hence its crystallographic morphology will also be different. The predicted morphology is shown in Figure S2b. The crystal habit of $Co_3Se_4$ is a rectangular box-like shape with longitudinal side— (100) and (-100) faces strongly visible, followed by (-201), (20-1) (001), and (00-1) planes. The other planes (110), (-110), (-1-10), (1-10) (1-1-1), (-1-11), (11-1), and (-111) are on the edges.

Since $Ni_3Se_4$ and $Fe_3Se_4$ have almost identical crystal properties, their crystallographic morphologies are likewise similar (as shown Figure S2c).

This model, however, has ignored one of the most essential factors in determining the crystal's morphology— bond energies and their orientations present in the crystal system. Therefore, it is crucial to discuss the proper consideration of several interaction energies present between the crystallizing entities in $M_3Se_4$ compounds.

In 1955, P. Hartman and W.G. Perdok[4–6] developed a model based on bond energies to predict any crystal's morphology. They demonstrated that the periodic chains of strong bonds present in the internal crystal structure play an essential role in deciding the crystal's final morphology. All the crystal faces are divided into three parts depending on the periodic chains of strong bonds named as PBC (periodic bond chain) vectors— F: flat faces (parallel to two PBC vectors), S: stepped faces (parallel to only one PBC vector), and K: kinked faces (not parallel to any PBC vector). The sequence of MI faces for appearance in the final crystal habit is F > S > K. Further, to predict the crystal morphology using this model, the inversion symmetry operator should be present. As if the inversion operator is not present, an intrinsic dipole moment is present within the unit cell, and this model works only for crystals with non-polar unit cells. The list of symmetry operators of $M_3Se_4$ is discussed in Figure S3. Given that all $M_3Se_4$ compounds contain the inversion operator, the HP model can be used to predict the crystal habit of these phases.

Figure S4 to S6 show the packing diagram of $M_3Se_4$ crystal having a blue cloud shape PBC vector. PBC vector completely lies on the (010) and (0-10) planes of the $M_3Se_4$ structure. Accordingly, these planes are in the F category of faces and are most visible in the final crystal shape in all three compounds. In case of $Fe_3Se_4$ and $Ni_3Se_4$, large component of PBC vectors lie on the (001), (00-1), (101), (-101), (10-1), (-10-1), (110), (-110), (1-10), (-1-10), (011), (0-11), (01-1), (0-1-1), (111), (-1-1-1), (-111), (1-11), (11-1), (1-1-1), (-11-1), and (-1-11) planes. While, in $Co_3Se_4$, large component of PBC vectors lie on the (100), (-100), (-201), (20-1), (101), (-10-1), (10-1), (-101), (110), (-110), (1-10), (-1-10), (011), (01-1), (0-11), (0-1-1), (111), (-1-1-1), (-111), (1-11), (11-1), (1-1-1), (-11-1), (1-1-1) planes. Therefore, these planes correspond to the S category of faces for the corresponding samples. In the case of $Fe_3Se_4$ and $Ni_3Se_4$, the (100) and (-100) planes are not parallel to any PBC vector, while, in $Co_3Se_4$, they are (001), (00-1) planes. Therefore, they belong to the K category having the least visibility in the final morphology. A table summarizing all $M_3Se_4$ crystal planes of different categories based on the HP model is Table S3— which provides the RGR and MI for distinct faces of corresponding $M_3Se_4$ compounds. Again, the WinXmorph[16] software has been used to estimate the crystal habit of $M_3Se_4$ compounds using these RGR and MI. It is observed that the (010) and (0-10) faces are of the highest MI for all $M_3Se_4$ crystals. In Figure S7, the morphological drawing predicted using HP theory illustrates the resultant crystal habit as a hexagonal-cylindrical-like shape with two enormously visible upper planes (010) and (0-10). It is noted that in all the $M_3Se_4$ compounds, the predicted crystal habit using the HP model is nearly the same with slight variation in $Co_3Se_4$ as bonding is nearly identical in all compounds.

**Table S1.** Rietveld refinement data of the as-synthesized $M_3Se_4$ nanoparticles —$Fe_3Se_4$, $Co_3Se_4$, and $Ni_3Se_4$. The parameters χ2 represented the squared ratio between R-factor from the refinement and R-factor expected from counting statistics, and $wR_P$ (%) is the weighted profile R-value. Unit cell parameters are denoted by a, b, c, α, β, and γ.[17]

| Compound / Parameter | $Fe_3Se_4$ | $Co_3Se_4$ | $Ni_3Se_4$ |
|---|---|---|---|
| $\chi^2$ | 1.69 | 15.2 | 24.8 |
| $wR_P$ (%) | 25.9 | 28.2 | 20.3 |
| Space group | I2/m | C2/m | I2/m |
| Structure | Monoclinic | Monoclinic | Monoclinic |
| a (Å) | 6.2 | 12.2 | 6.2 |
| b (Å) | 3.5 | 3.5 | 3.6 |
| c (Å) | 11.3 | 6.2 | 10.5 |
| α | 90° | 90° | 90° |
| β | 91.8° | 121.8° | 90.7° |
| γ | 90° | 90° | 90° |

**Table S2**. Morphological importance of several faces of $M_3Se_4$ crystals ($Fe_3Se_4$, $Co_3Se_4$, and $Ni_3Se_4$) based on the Bravais-Friedel-Donnay-Harker (BFDH) model. The $d_{hkl}$ represents the interplanar spacing in Å. RGR and MI are calculated as 'relative growth rates' and 'morphological importance', respectively.

| Phase | Faces (hkl) | $d_{hkl}$ (Å) | RGR | MI |
|---|---|---|---|---|
| $Fe_3Se_4$ | $(001), (00\bar{1})$ | 11.27 | 1 | 1 |
| | $(\bar{1}01), (10\bar{1})$ | 5.51 | 2.04 | 0.49 |
| | $(101), (\bar{1}0\bar{1})$ | 5.36 | 2.10 | 0.48 |
| | $(011), (0\bar{1}1), (0\bar{1}\bar{1}), (01\bar{1})$ | 3.37 | 3.34 | 0.29 |
| | $(110), (\bar{1}10), (\bar{1}\bar{1}0), (1\bar{1}0)$ | 3.07 | 3.66 | 0.27 |
| $Co_3Se_4$ | $(100), (\bar{1}00)$ | 10.40 | 1 | 1 |
| | $(001), (00\bar{1})$ | 5.34 | 1.94 | 0.51 |
| | $(\bar{2}01), (20\bar{1})$ | 5.21 | 1.99 | 0.49 |
| | $(110), (\bar{1}10), (\bar{1}\bar{1}0), (1\bar{1}0)$ | 3.37 | 3.07 | 0.32 |
| | $(1\bar{1}\bar{1}), (\bar{1}\bar{1}1), (11\bar{1}), (\bar{1}11)$ | 3.08 | 3.36 | 0.29 |
| $Ni_3Se_4$ | $(001), (00\bar{1})$ | 10.25 | 1 | 1 |
| | $(\bar{1}01), (10\bar{1}), (101), (\bar{1}0\bar{1})$ | 5.32 | 1.96 | 0.50 |
| | $(011), (0\bar{1}1), (0\bar{1}\bar{1}), (01\bar{1})$ | 3.43 | 3.04 | 0.32 |
| | $(110), (\bar{1}10), (\bar{1}\bar{1}0), (1\bar{1}0)$ | 3.13 | 3.33 | 0.29 |

**Table S3**. Morphological importance of several faces of $M_3Se_4$ ($Fe_3Se_4$, $Co_3Se_4$, and $Ni_3Se_4$) crystals based on the Hartman-Perdok (HP) model. All the faces are categorized according to their alignment with PBC vectors. Here, MI stands for 'morphological importance'.

| S. No. | Faces (hkl) | PBC vectors lying parallel to (hkl) planes | Face category | MI |
|---|---|---|---|---|
| $Fe_3Se_4$ and $Ni_3Se_4$ | $(010)$, $(0\bar{1}0)$ | Blue | F | Largest |
| | $(001)$, $(00\bar{1})$,<br>$(101)$, $(\bar{1}0\bar{1})$, $(10\bar{1})$, $(\bar{1}01)$,<br>$(110)$, $(\bar{1}10)$, $(1\bar{1}0)$, $(\bar{1}\bar{1}0)$,<br>$(011)$, $(0\bar{1}1)$, $(01\bar{1})$, $(0\bar{1}\bar{1})$,<br>$(111)$, $(\bar{1}\bar{1}\bar{1})$, $(\bar{1}11)$, $(1\bar{1}1)$,<br>$(11\bar{1})$, $(1\bar{1}\bar{1})$, $(\bar{1}1\bar{1})$, $(\bar{1}\bar{1}1)$ | Large component of blue | S | Large |
| | $(100)$, $(\bar{1}00)$ | No PBC vector lie along these planes | K | Smallest |
| $Co_3Se_4$ | $(010)$, $(0\bar{1}0)$ | Blue | F | Largest |
| | $(100)$, $(\bar{1}00)$, $(\bar{2}01)$, $(20\bar{1})$,<br>$(101)$, $(\bar{1}0\bar{1})$, $(10\bar{1})$, $(\bar{1}01)$,<br>$(110)$, $(\bar{1}10)$, $(1\bar{1}0)$, $(\bar{1}\bar{1}0)$,<br>$(011)$, $(0\bar{1}1)$, $(01\bar{1})$, $(0\bar{1}\bar{1})$,<br>$(111)$, $(\bar{1}\bar{1}\bar{1})$, $(\bar{1}11)$, $(1\bar{1}1)$,<br>$(11\bar{1})$, $(1\bar{1}\bar{1})$, $(\bar{1}1\bar{1})$, $(\bar{1}\bar{1}1)$ | Large component of blue | S | Large |
| | $(001)$, $(00\bar{1})$ | No PBC vector lie along these planes | K | Smallest |

**Table S4.** Comparison of magnetic properties of $M_3X_4$ compounds (where M can be Fe, Co, and Ni, and X can be O, S, and Se) with their corresponding magnetic transition temperature.

| ($M_3X_4$) | | | |
|---|---|---|---|
| X \ M | **Fe** | **Co** | **Ni** |
| **O** | Ferrimagnetic[18–22] ($T_C$ ~ 858 K) | Antiferromagnetic[23] ($T_N$ ~ 40 K) | Ferrimagnetic[24–28] ($T_C$ ~ 808 K) |
| **S** | Ferrimagnetic[29–34] ($T_C$ ~ 606 K) | Paramagnetic[35]/ Antiferromagnetic[36,37] ($T_N$ ~ 58 K) | Ferrimagnetic[38,39] ($T_C$ ~ 20 K) |
| **Se** | Ferrimagnetic[40,41] ($T_C$ ~ 314 K) | Paramagnetic[42] | Paramagnetic[42] |

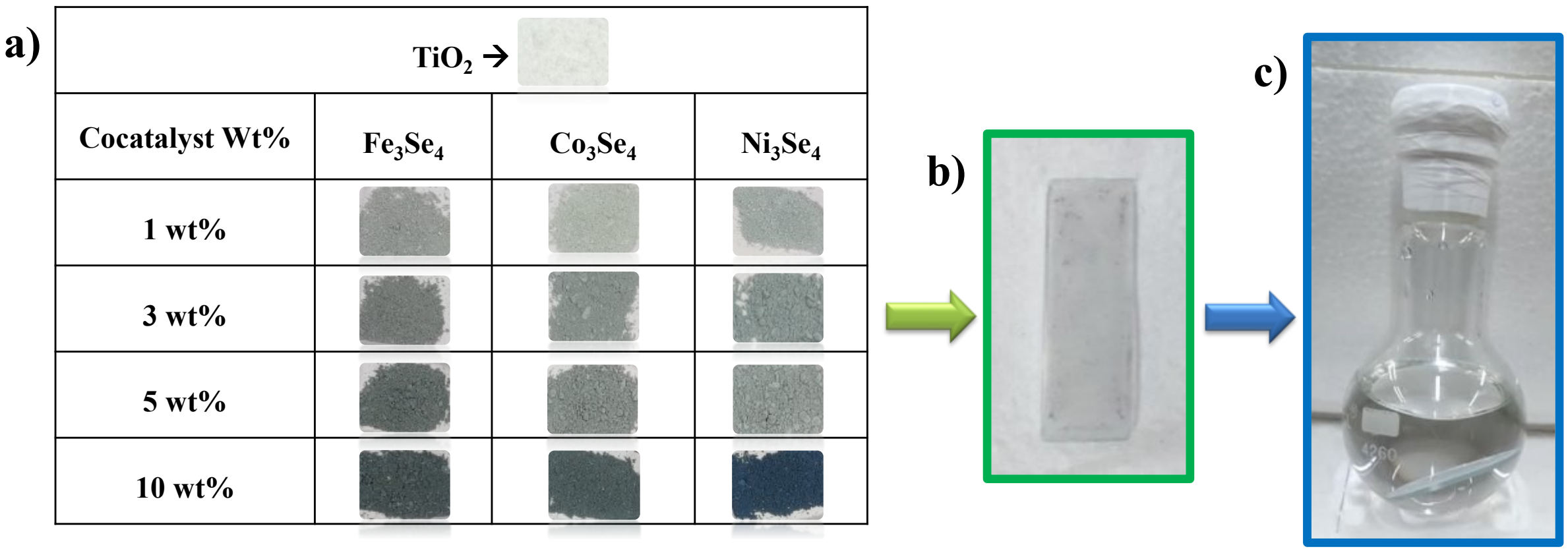

**Figure S1**. a) Tabulated the colour of all distinct sample, b) photographs of thin film which is used for hydrogen production using round bottom flask setup as shown in c).

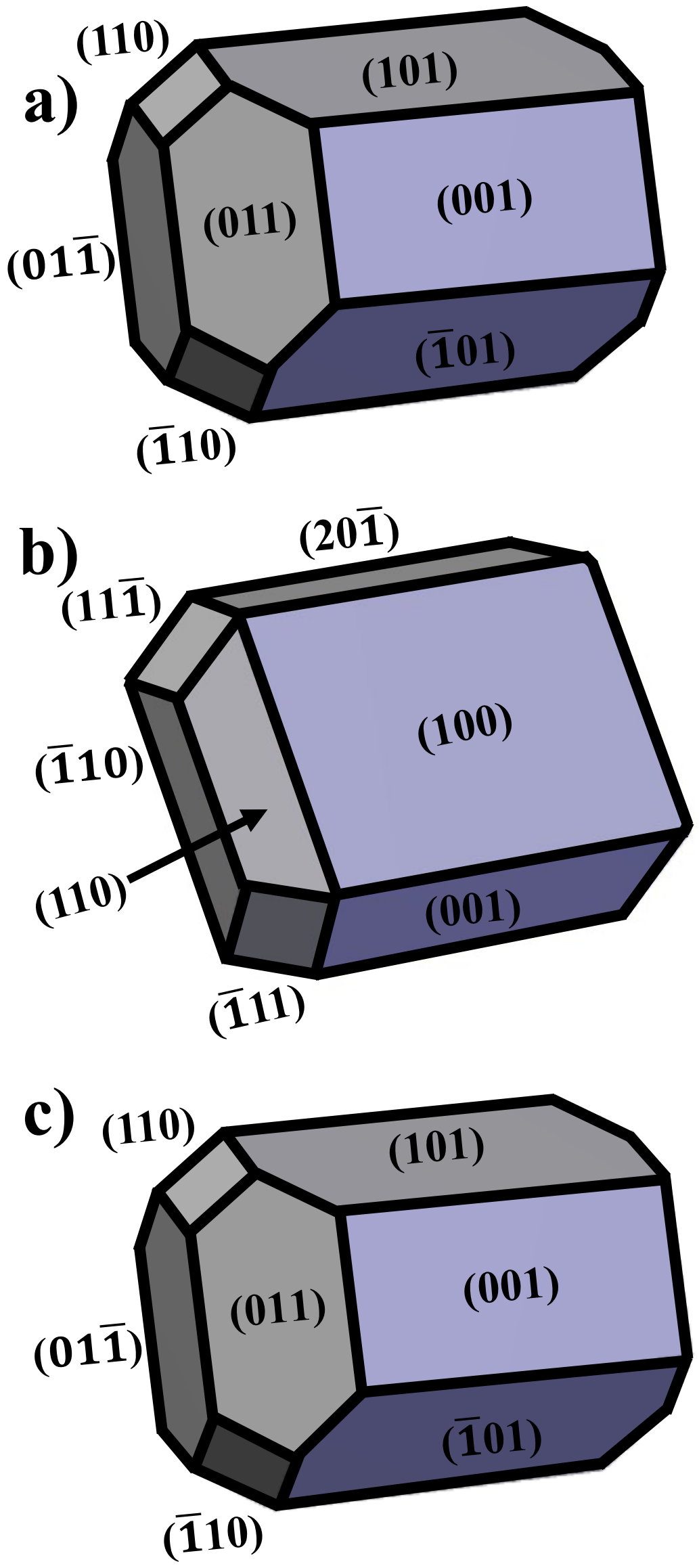

**Figure S2.** The predicted morphologies of $M_3Se_4$ crystals using Bravais-Friedel-Donnay-Harker (BFDH) model. a) $Fe_3Se_4$, b) $Co_3Se_4$, and c) $Ni_3Se_4$ show the indexed morphological drawing with corresponding hkl planes.

**$Fe_3Se_4$**

| Symm. Op. | Description |
|---|---|
| x,y,z | Identity |
| -x,y,-z | Rotation axis (2-fold) |
| -x,-y,-z | Inversion centre |
| x,-y,z | Mirror plane |
| 1/2+x,1/2+y,1/2+z | Centring vector |
| 1/2-x,1/2+y,1/2-z | Screw axis (2-fold) |
| 1/2-x,1/2-y,1/2-z | Inversion centre |
| 1/2+x,1/2-y,1/2+z | Glide plane |

**$Co_3Se_4$**

| Symm. Op. | Description |
|---|---|
| x,y,z | Identity |
| 1/2+x,1/2+y,z | Centring vector |
| x,-y,z | Mirror plane |
| 1/2+x,1/2-y,z | Glide plane |
| -x,y,-z | Rotation axis (2-fold) |
| 1/2-x,1/2+y,-z | Screw axis (2-fold) |
| -x,-y,-z | Inversion centre |
| 1/2-x,1/2-y,-z | Inversion centre |

**$Ni_3Se_4$**

| Symm. Op. | Description |
|---|---|
| x,y,z | Identity |
| 1/2+x,1/2+y,1/2+z | Centring vector |
| x,-y,z | Mirror plane |
| 1/2+x,1/2-y,1/2+z | Glide plane |
| -x,y,-z | Rotation axis (2-fold) |
| 1/2-x,1/2+y,1/2-z | Screw axis (2-fold) |
| -x,-y,-z | Inversion centre |
| 1/2-x,1/2-y,1/2-z | Inversion centre |

**Figure S3**. The list of symmetry operators of $M_3Se_4$. a) $Fe_3Se_4$, b) $Co_3Se_4$, and c) $Ni_3Se_4$.

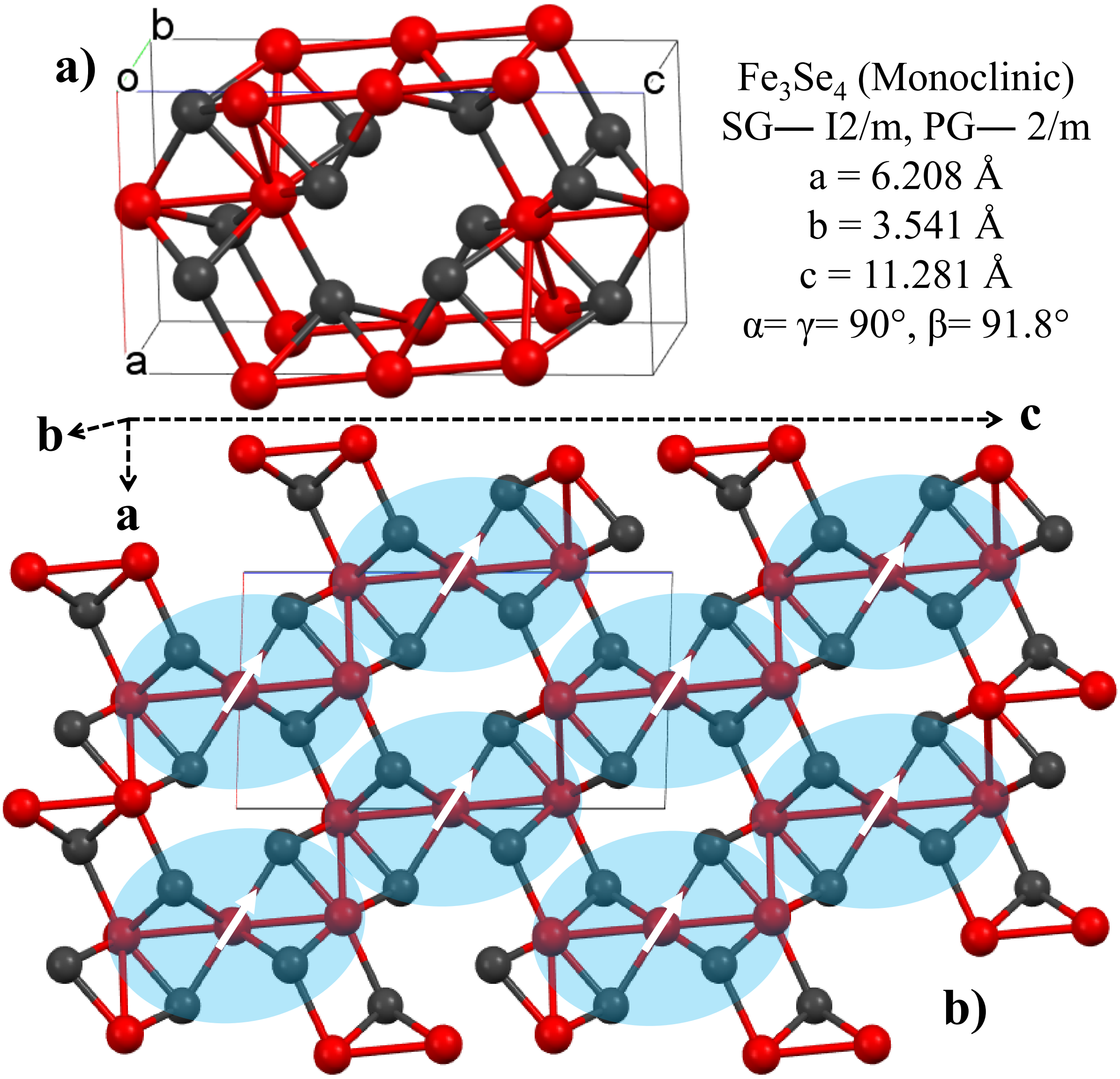

**Figure S4.** a) Unit cell of $Fe_3Se_4$. b) Packing diagram of $Fe_3Se_4$ crystal along with 1 PBC vector represented by blue cloud having stoichiometry $Fe_3Se_4$. The three crystallographic axes are denoted by a, b, and c.

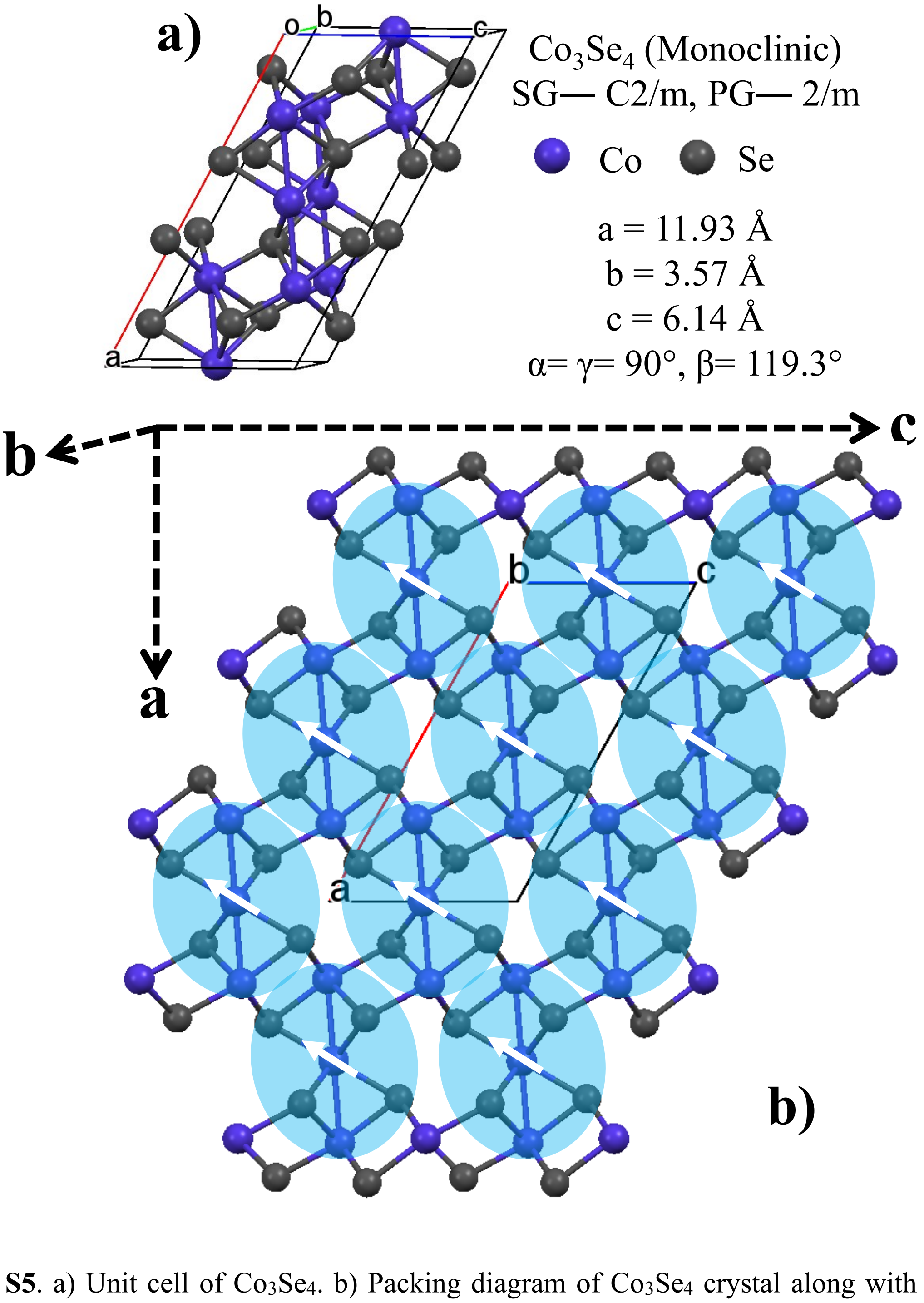

**Figure S5**. a) Unit cell of $Co_3Se_4$. b) Packing diagram of $Co_3Se_4$ crystal along with 1 PBC vector represented by blue cloud having stoichiometry $Co_3Se_4$. The three crystallographic axes are denoted by a, b, and c.

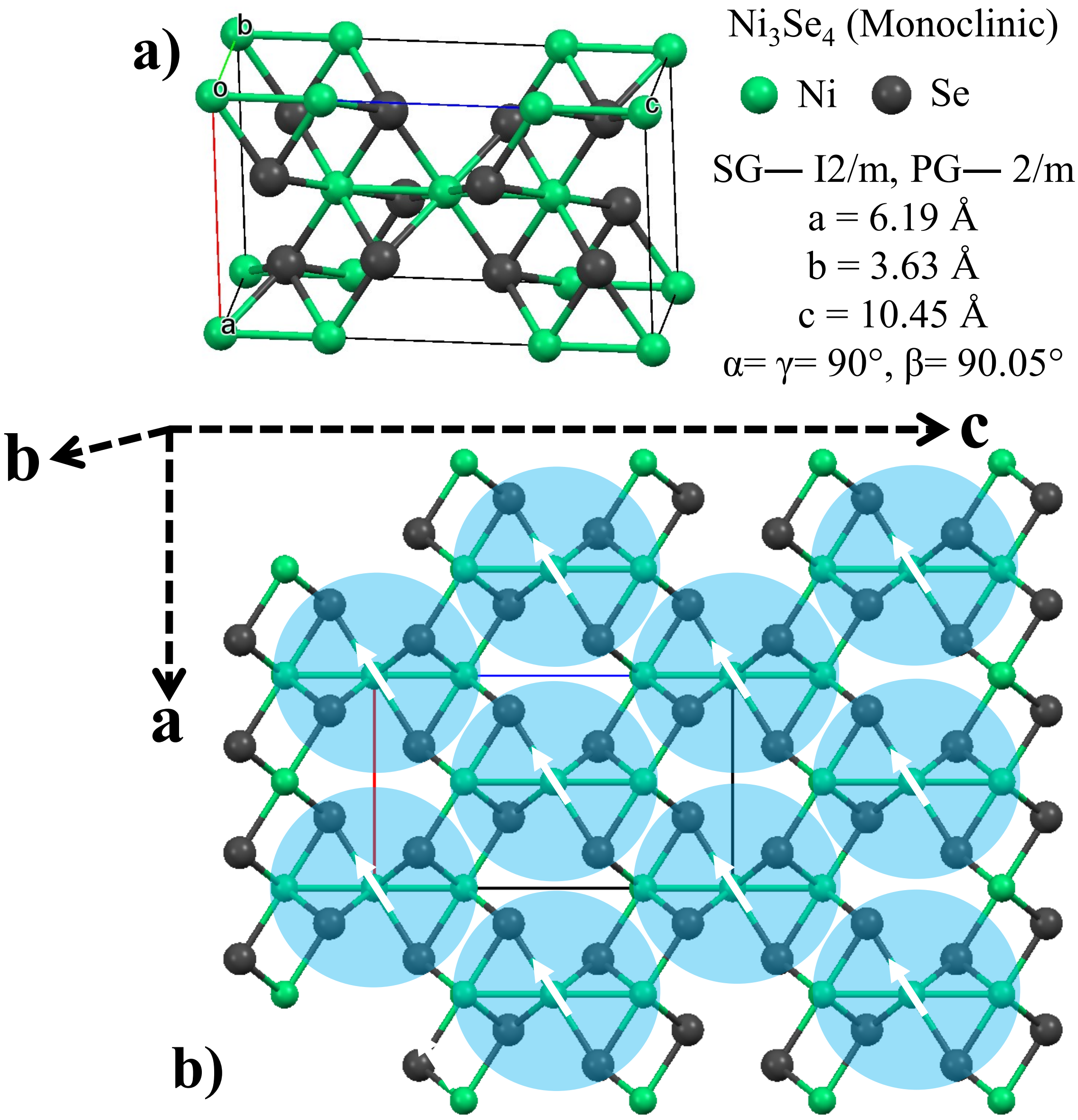

**Figure S6.** a) Unit cell of $Ni_3Se_4$. b) Packing diagram of $Ni_3Se_4$ crystal along with 1 PBC vector represented by blue cloud having stoichiometry $Ni_3Se_4$. The three crystallographic axes are denoted by a, b, and c.

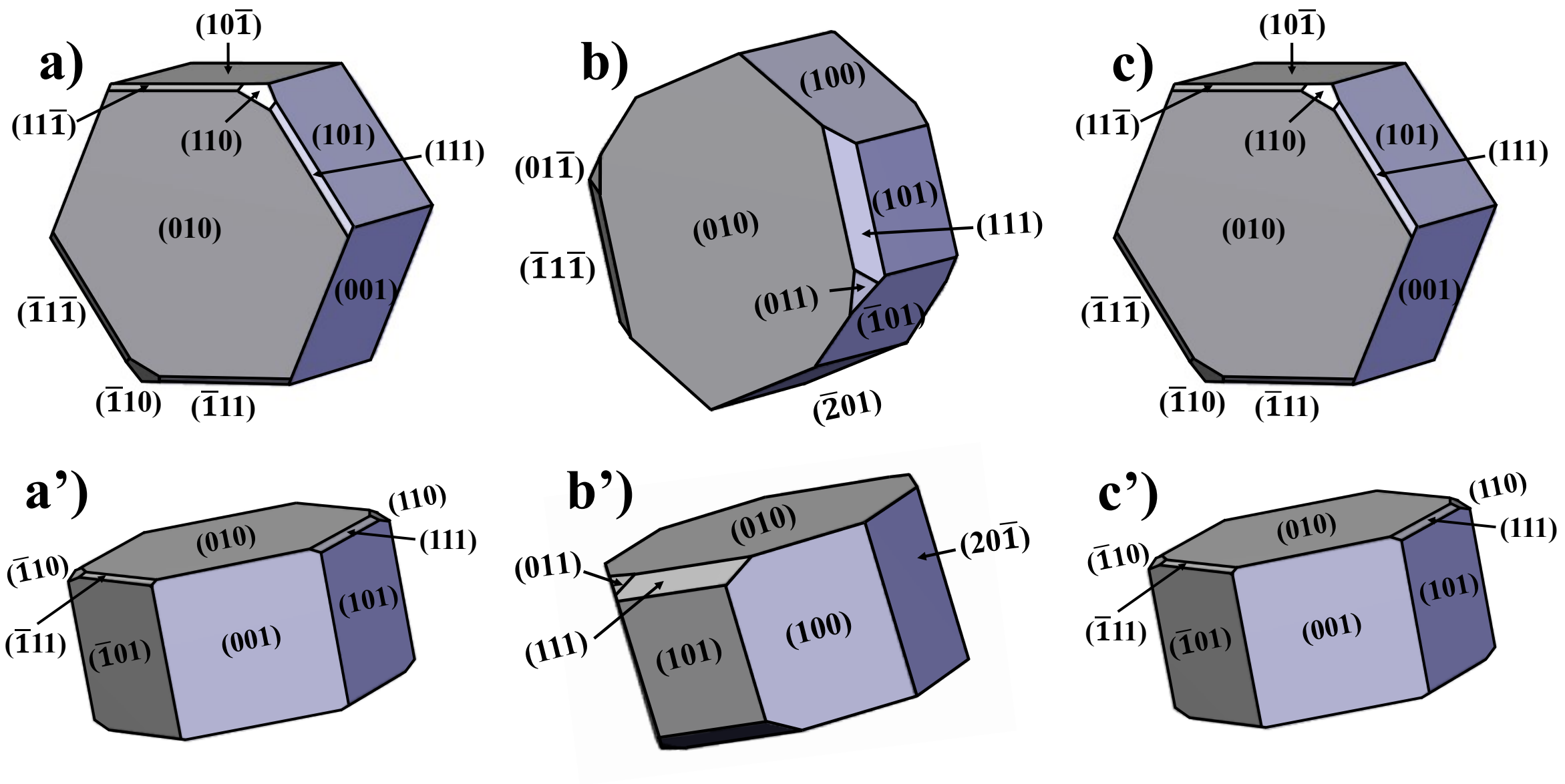

**Figure S7.** Calculated crystal growth morphologies of $M_3Se_4$ crystals predicted using Hartman-Perdok (HP) model with corresponding hkl planes. Figures a) to c) show the top view (in-plane) of the morphologies of a) $Fe_3Se_4$, b) $Co_3Se_4$, and c) $Ni_3Se_4$ compounds with the highest morphological importance of the (010) plane. Figures a') to c') show the side-views of the morphologies of corresponding compounds.

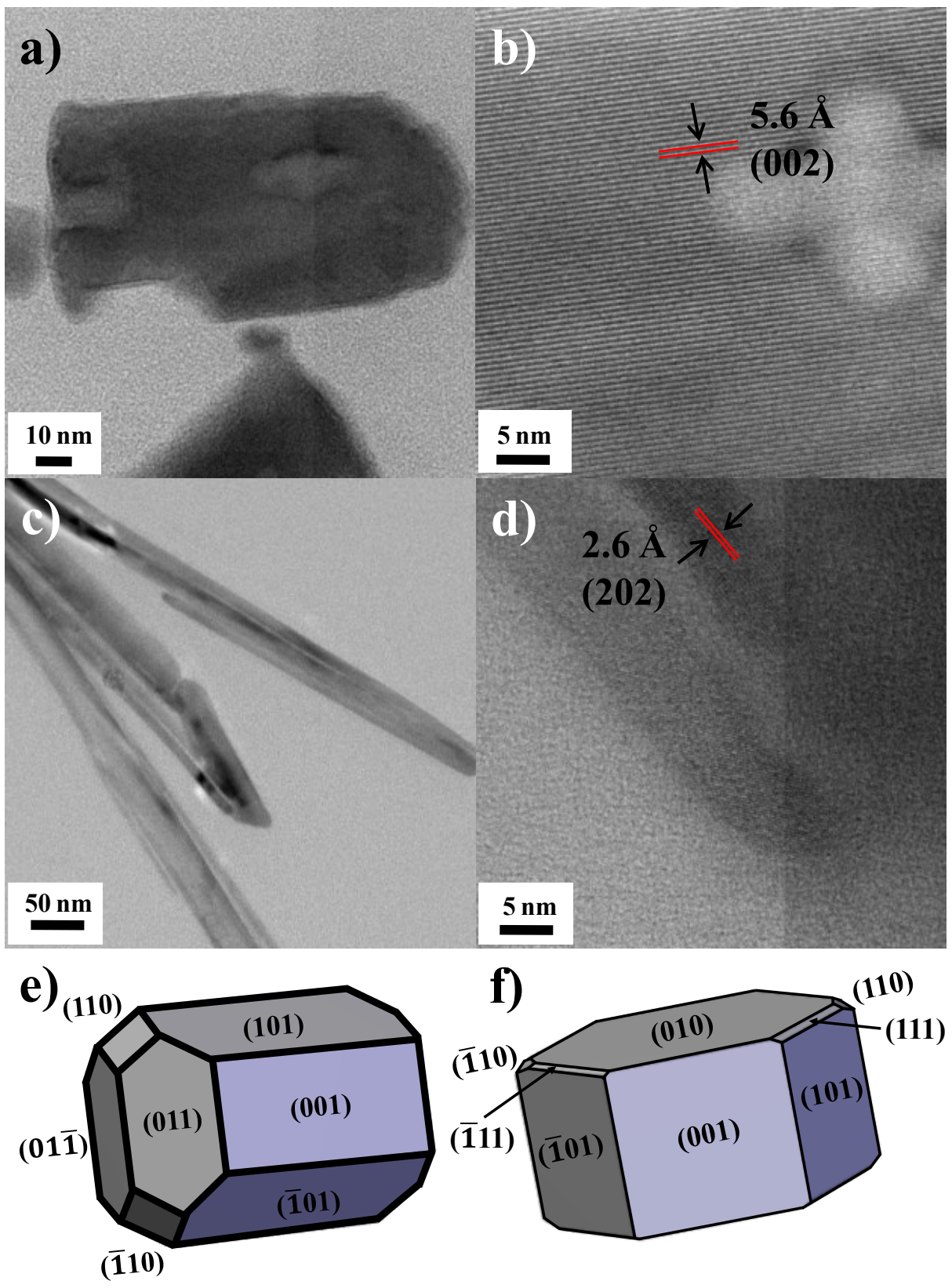

**Figure S8.** Correlation between the experimental and theoretically predicted morphologies of $Fe_3Se_4$. Figures a) and c) represent the TEM images of $Fe_3Se_4$ nanoparticles having the rod-like growth. Figures b) and d) show the lattice fringes spaced at 5.6 Å, and 2.6 Å representing the (002) and (202) planes of $Fe_3Se_4$ respectively.[29] Figures e) and f) represent the predicted morphology of $Fe_3Se_4$ by BFDH and HP model, respectively.

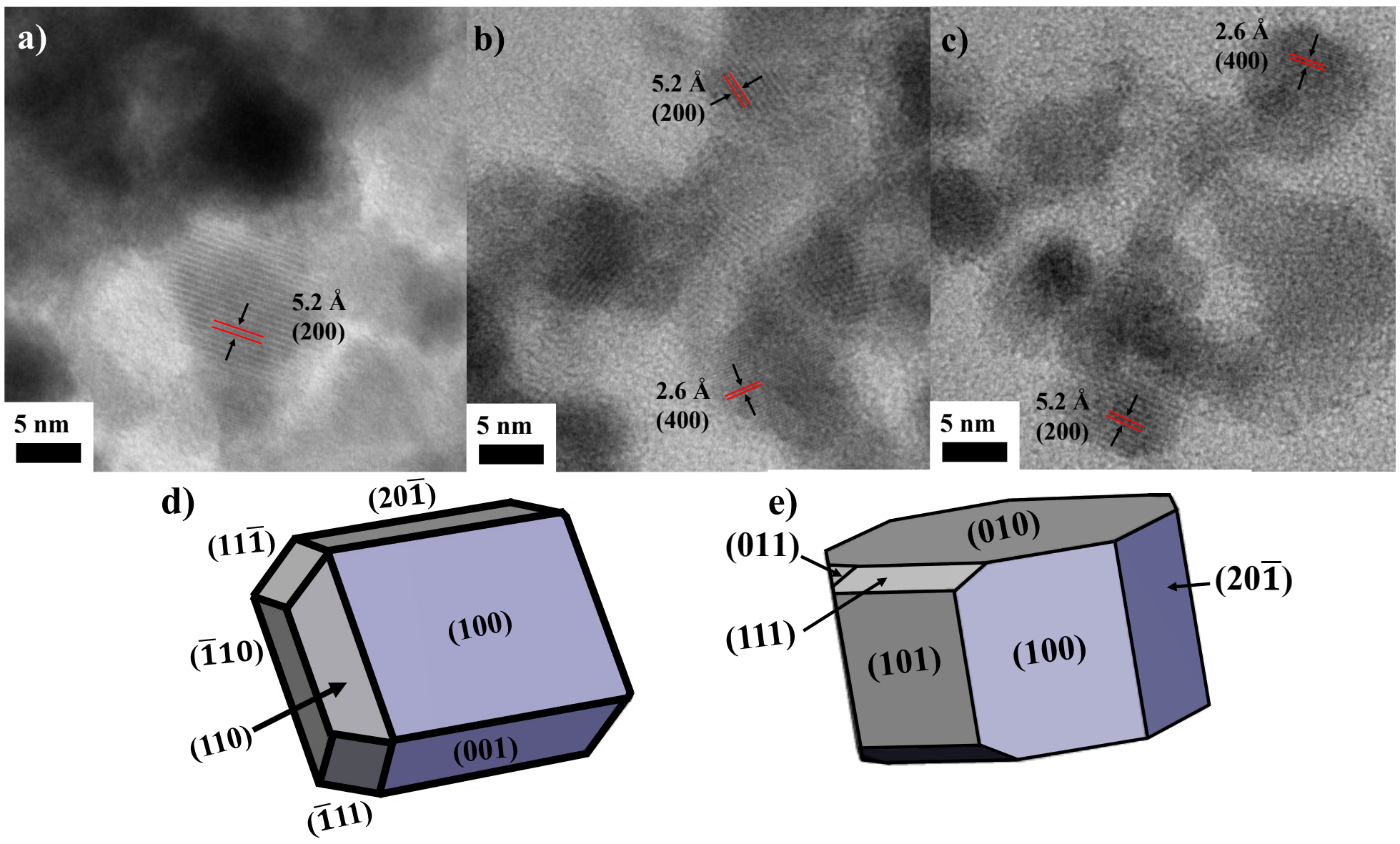

**Figure S9.** Correlation between experimentally-observed and theoretically-predicted morphologies of $Co_3Se_4$. Figures a) to c) represent the TEM images of $Co_3Se_4$ nanoparticles having quasi-spherical features. Figures a), b), and c) also show the lattice fringes spaced at 5.2 Å and 2.6 Å, which represent the (200) and (400) planes of $Co_3Se_4$, respectively. Figures d) and e) represent the predicted morphology of $Co_3Se_4$ by BFDH and HP model, respectively.

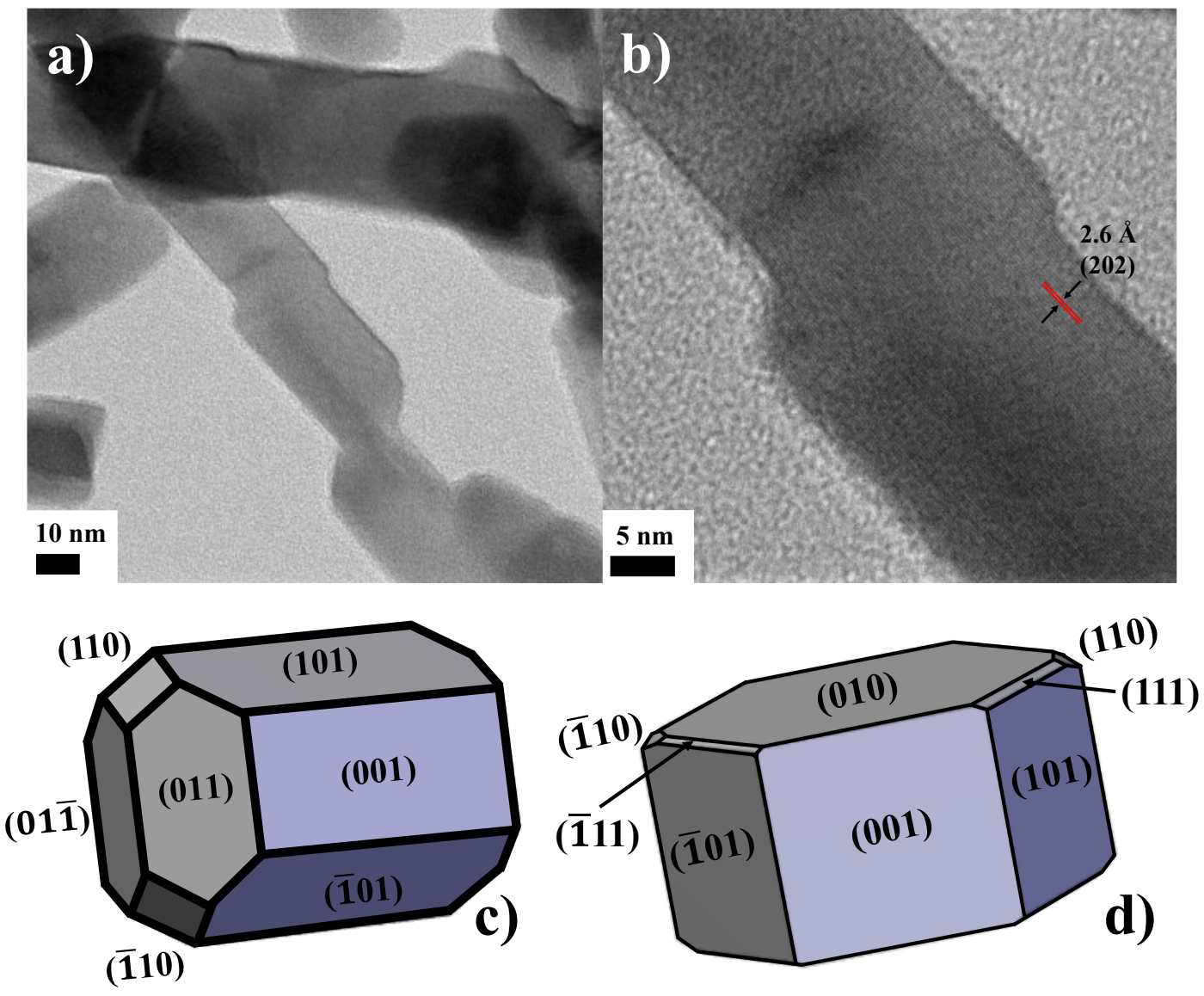

**Figure S10.** Correlation between the experimentally-observed and theoretically-predicted morphologies of $Ni_3Se_4$. Figures a) and b) represent the TEM images of $Ni_3Se_4$ nanoparticles with rod-like features. Figure b) shows the lattice fringes spaced at 2.6 Å, representing the (202) plane of $Ni_3Se_4$. Figures c) and d) represent the predicted morphology of $Ni_3Se_4$ by BFDH and HP model, respectively.

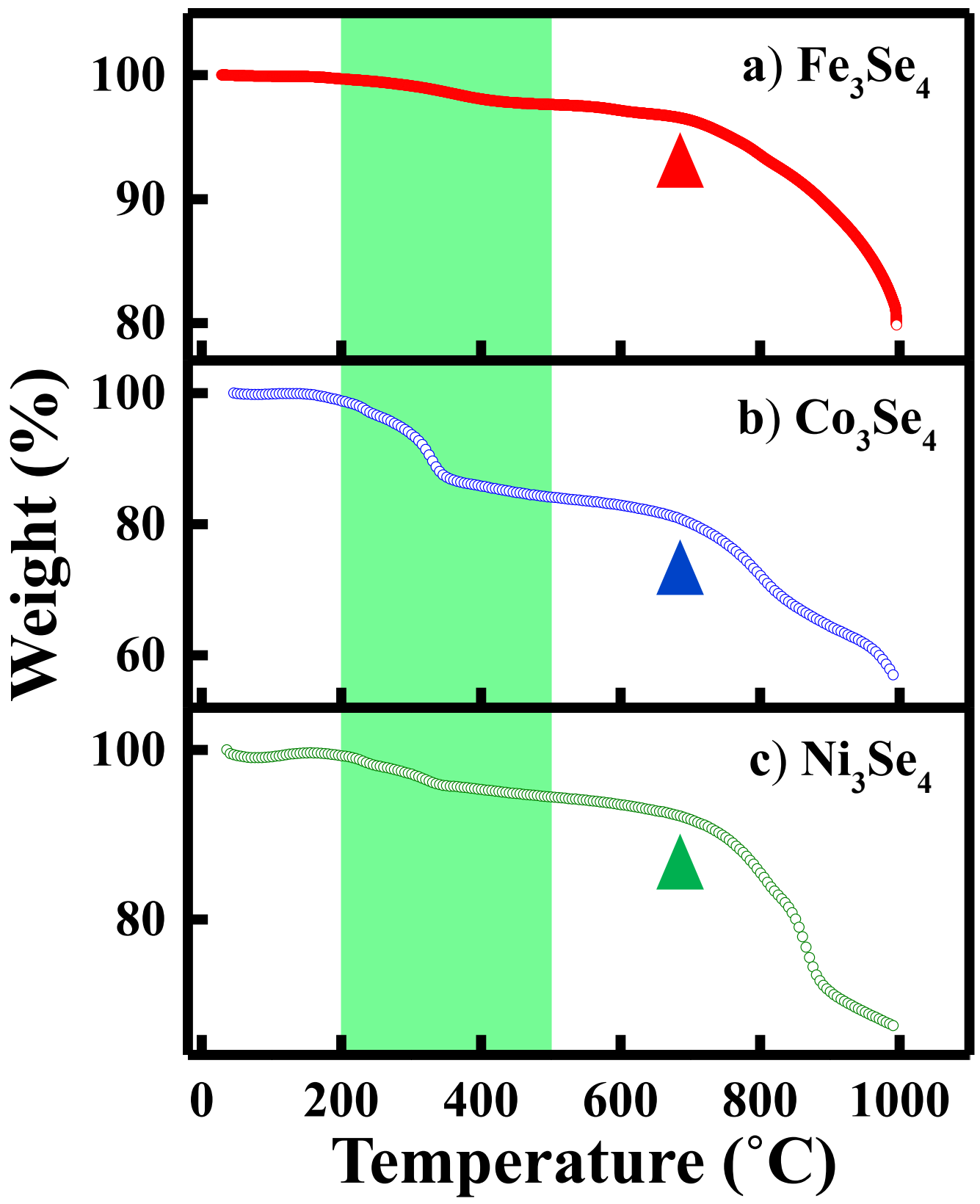

**Figure S11.** TGA plots of a) $Fe_3Se_4$[25] (red curve), b) $Co_3Se_4$ (blue curve), and c) $Ni_3Se_4$ (green curve). The shaded region shows the loss of organic fragments, and the triangle reveals the decomposition point in corresponding compounds.

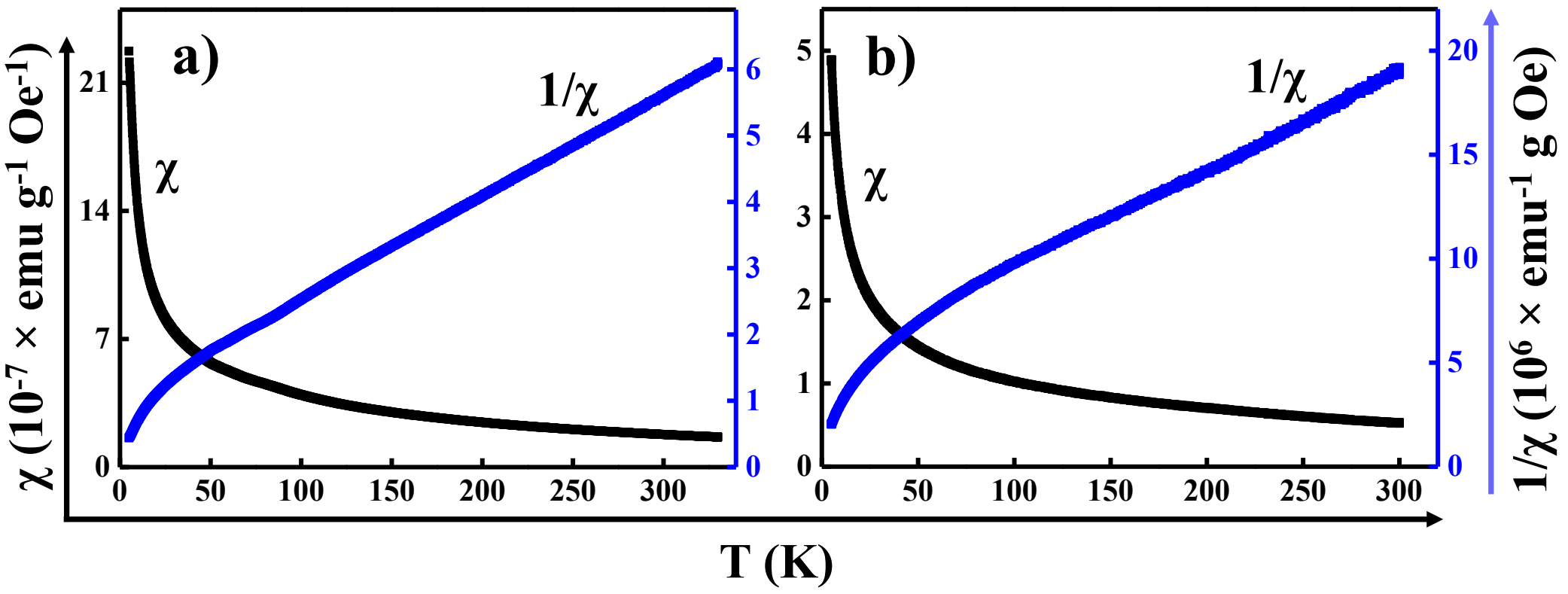

**Figure S12.** The black curve represents the magnetic susceptibility (χ) versus the temperature plot of a) $Co_3Se_4$ and b) $Ni_3Se_4$, respectively, at 100 Oe. The blue curve denotes the inverse of the magnetic susceptibility versus temperature plot.

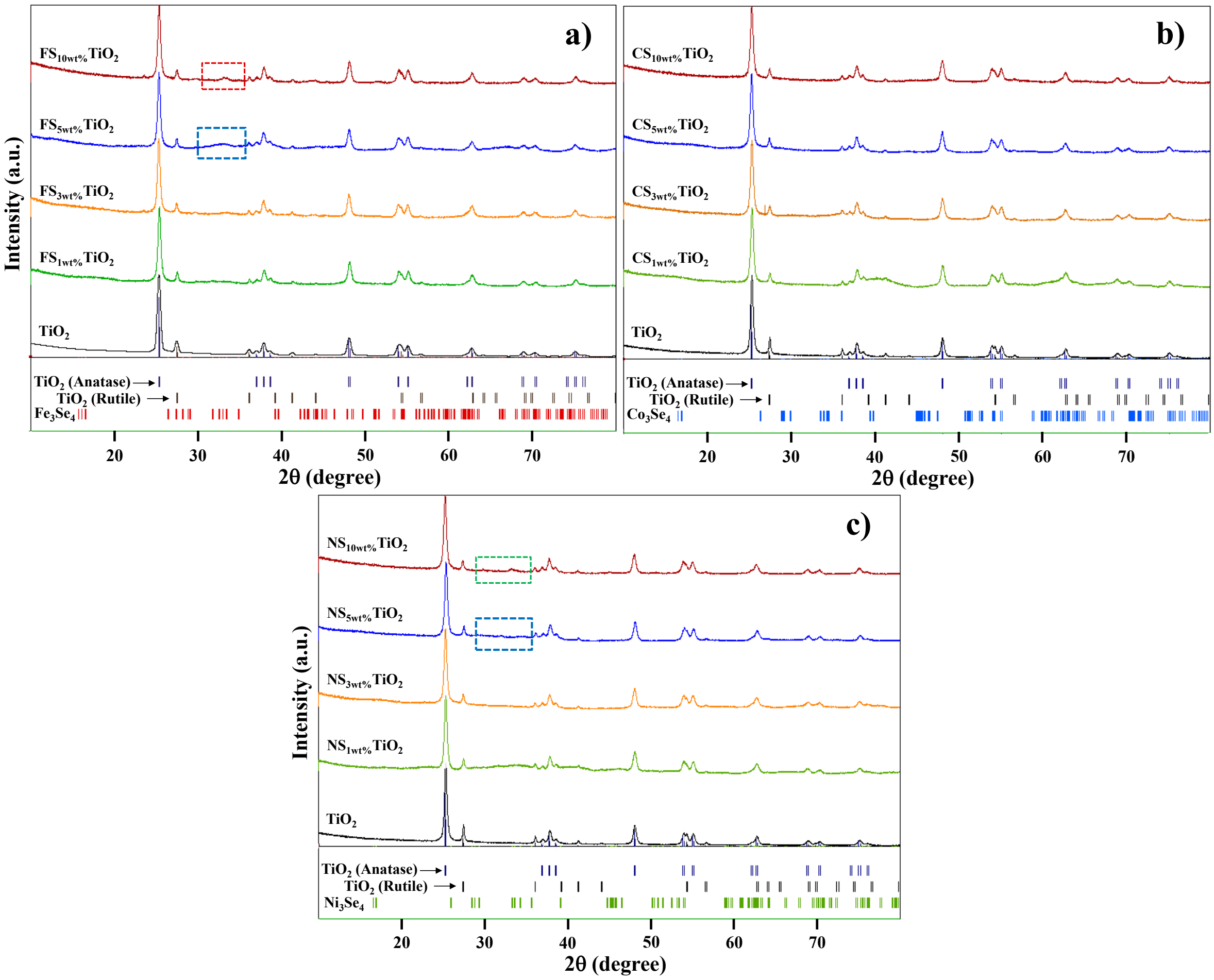

**Figure S13**. XRD patterns of $M_3Se_4$ $_{(1\ to\ 10\ wt\%)}$-$TiO_2$ with different loading amounts of $M_3Se_4$.

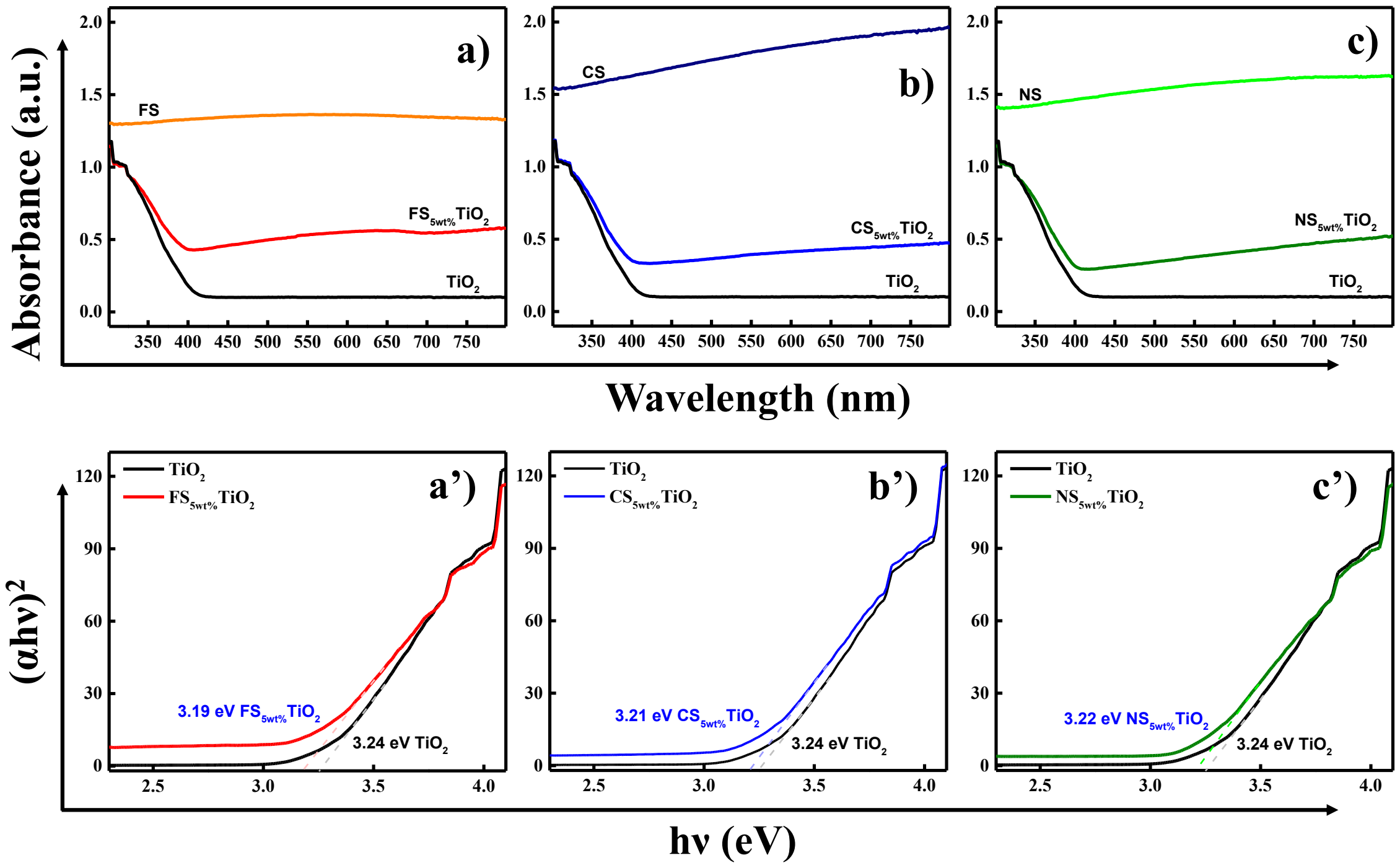

**Figure S14**. (a-c) UV-Visible absorption spectra of $TiO_2$, $M_3Se_4$ and $M_3Se_4$ $_{(5wt\%)}$-$TiO_2$,; a)$Fe_3Se_4$ (FS), b) $Co_3Se_4$(CS), and c)$Ni_3Se_4$(NS). (b) Tauc plot of $TiO_2$ and $M_3Se_4$ $_{(5wt\%)}$-$TiO_2$,; a')$Fe_3Se_4$ (FS) $_{(5wt\%)}$-$TiO_2$, b') $Co_3Se_4$ $_{(5wt\%)}$-$TiO_2$, and c')$Ni_3Se_4$$_{(5wt\%)}$-$TiO_2$ for their respective band gap indicated by arrows.